\documentclass[12pt,preprint]{aastex}

\newcommand{\msun}{{~{\rm M}_\odot}}

\newcommand{\etal}{{et~al.}~}

\shorttitle{}
\shortauthors{Hicks et al.}

\begin{document}

\title{Multiwavelength Mass Comparisons of the z$\sim$0.3 CNOC Cluster Sample}

\author{A.K. Hicks\email{ahicks@alum.mit.edu}
\affil{Department of Astronomy, University of Virginia, P.O. Box 400325,
Charlottesville, VA 22904}} 

\author{E. Ellingson\email{elling@casa.colorado.edu} 
\affil{Center for Astrophysics and Space Astronomy, University of Colorado at Boulder, Campus Box 389, Boulder, CO 80309}}

\author{H. Hoekstra\email{hoekstra@uvic.ca}}
\affil{Department of Physics \& Astronomy, University of Victoria, Elliott Building, 3800 Finnerty Rd, Victoria, BC, V8P 5C2}

\and

\author{H.K.C. Yee\email{hyee@astro.utoronto.ca}}
\affil{Department of Astronomy and Astrophysics, University of Toronto, 60 St. George St., Toronto, ON, M5S 3H8, Canada}

\begin{abstract}
Results are presented from a detailed analysis of optical and X-ray observations of moderate-redshift galaxy clusters from the Canadian Network for Observational Cosmology (CNOC) subsample of the EMSS.  The combination of extensive optical and deep X-ray observations of these clusters make them ideal candidates for multiwavelength mass comparison studies.  X-ray surface brightness profiles of 14 clusters with $0.17<z<0.55$ are constructed from Chandra observations and fit to single and double $\beta$ models.  Spatially resolved temperature analysis is performed, indicating that five of the clusters in this sample exhibit temperature gradients within their inner $60-200$ kpc.  Integrated spectra extracted within $\rm{R}_{2500}$ provide temperature, abundance, and luminosity information.  Under assumptions of hydrostatic equilibrium and spherical symmetry, we derive gas and total masses within $\rm{R}_{2500}$ and $\rm{R}_{200}$.  We find an average gas mass fraction of $f_{gas}(\rm{R}_{200})=0.136\pm{0.004}~h_{70}^{-3/2}$, resulting in $\Omega_m=0.28\pm{0.01}$ (formal error).  We also derive dynamical masses for these clusters to $\rm{R}_{200}$.  We find no systematic bias between X-ray and dynamical methods across the sample, with an average $\rm{M}_{dyn}/\rm{M}_{X-ray} = 0.97\pm{0.05}$.  We also compare X-ray masses to weak lensing mass estimates of a subset of our sample, resulting in a weighted average of $\rm{M}_{lens}/\rm{M}_{X-ray}$ of $0.99\pm{0.07}$.  We investigate X-ray scaling relationships and find powerlaw slopes which are slightly steeper than the predictions of self-similar models, with an $E(z)^{-1}$ $L_X$-$T_X$ slope of $2.4\pm{0.2}$ and an $E(z)~\rm{M}_{2500}$-$T_X$ slope of $1.7\pm{0.1}$.  Relationships between red-sequence optical richness ($\rm{B}_{gc,red}$) and global cluster X-ray properties ($T_X$, $L_X$ and $\rm{M}_{2500}$) are also examined and fitted.

\end{abstract}


\section{Introduction \label{s:intro}}

Clusters of galaxies are valuable cosmological probes.  As the largest gravitationally bound objects in the universe, they play a key role in the tracing and modeling of large scale structure formation and evolution \citep[e.g.,][]{voit,bahcall}.  Because of our rapidly expanding sample of known clusters, finding efficient means of estimating cluster properties is a highly desirable goal.

The contribution of cluster studies to the field of observational cosmology hinges on our ability to accurately estimate cluster masses.  In particular, through the determination of both gas mass and total mass, cluster analysis can lead to estimations of the cosmological mass density, $\Omega_m$, while accurate measurement of the evolution of the cluster mass function provides important constraints on both the normalization of the matter power spectrum, $\sigma_8$ and $w$, the dark energy equation of state~\citep[e.g.,][]{levine02,eke}.  In addition, gaining an understanding of the evolution of X-ray scaling relationships (such as $L_X$-$T_X$ and M-$T_X$) with redshift provides an important contribution to our ability to accurately model the evolution of large-scale structure in the universe~\citep[e.g.,][]{voit}.  

Intermediate redshift ($0.2 < z < 0.6$) clusters are well situated for the study of cluster properties.  They are compact enough to be observed without tedious mosaicing, and they are present in statistically significant numbers.  Intermediate redshift clusters are also luminous enough to permit detailed investigation, with the potential for placing strong contraints on cluster evolution.

Three frequently applied approaches to estimating cluster masses are gravitational lens modeling, optical spectroscopic measurements of the cluster galaxy velocity dispersion, and characterizing X-ray emission as a means of tracing the underlying potential well of the cluster.  Each of these methods uses different observations and assumptions, which can be tested through their direct comparison.  An alternate optical approach to the efficient estimation of cluster masses involves the use of optical richness.  This method relies on the assumption that galaxy light is a reliable tracer of total cluster mass, and requires calibration via other methods.

Lensing mass estimates test both the assumption of hydrostatic equilibrium and our knowledge of the mass distribution in clusters.  Because they probe all of the projected mass along the line of sight, which may include additional mass concentrations, they are susceptible to overestimation of cluster mass~\citep{cen, metzler01}.  Dynamical mass estimates work under the assumption that velocity dispersion is directly related to the underlying gravitational potential of the cluster.  Pitfalls of this technique also include the danger of overestimating the cluster mass in cases where substructure or mergers drive up velocity dispersion measurements~\citep{bird}.  

The hot, diffuse intracluster medium (ICM), which is observed in the X-ray, should be a direct tracer of a cluster's underlying potential well.  Under assumptions of isothermality and hydrostatic equilibrium, the surface brightness of a cluster of galaxies can provide information on gas density as well as total gravitating mass.  Factors that influence the accuracy of X-ray mass estimates are temperature gradients, substructure, and mergers, which can compromise the previously stated assumptions~\citep[e.g.,][]{balland}.

One of the main objectives of this work is to identify, through direct comparison, any systematic biases in these methods with the ultimate goal of determining a robust calibration between optical richness and cluster mass.  Optical richness measurements are easily available due to the fact that their estimation requires very little observing time. Therefore an accurate calibration of the relationship between mass and optical richness would allow us to determine the masses of large samples of clusters in a highly efficient manner, providing strong constraints on the evolution of the cluster mass function and, consequently, key cosmological parameters.

In this paper we present a detailed analysis of high resolution Chandra X-ray observations of 14 CNOC~\citep{yee96} clusters at z $\sim 0.3$.  In Sections ~\ref{s:obs}-\ref{s:spect} we probe the temperature, metallicity, morphology, and surface brightness of the hot ICM present in each cluster.  From this initial analysis, we derive mass estimates (Section~\ref{s:mass}) which are compared to dynamical and weak lensing mass estimates (Section~\ref{s:masscomp}), and then we examine the X-ray scaling laws of our sample (Section~\ref{s:laws}).  Finally, in Section~\ref{s:or}, we use our results to calibrate relationships between red-sequence optical richness ($\rm{B}_{gc,red}$) and global cluster X-ray properties ($T_X$, $L_X$ and $\rm{M}_{2500}$).

Unless otherwise noted, this paper assumes a cosmology of $\rm{H}_0=70~\rm{km}~\rm{s}^{-1}~\rm{Mpc}^{-1}$, $\Omega_m=0.3$, and $\Omega_{\Lambda}=0.7$

\section{Cluster Sample \& Chandra Observations \label{s:obs}}

X-ray observations of our sample were retrieved from the Chandra Data Archive
(CDA) after conducting a search for currently available
Chandra observations of the Canadian Network for Observational Cosmology \citep[CNOC]{yee96} intermediate redshift ($0.17<z<0.55$) subsample of 15 EMSS clusters~\citep{gioia90,henry92} and one Abell cluster~\citep{abell}.  The selection criterion for this sample can be found in~\citet{yee96}. This sample has been extensively observed by the CNOC cluster survey (CNOC-1), and galaxy redshifts of $\sim 1200$ cluster members as well as detailed photometric catalogues are available for these clusters \citep[e.g.,][]{cnoccat}.

Chandra Advanced CCD Imaging Spectrometer (ACIS) observations of 14 of the 16 CNOC clusters were obtained.  Six of these clusters were observed with the ACIS-S CCD array, and eight were observed with ACIS-I, with an overall range in individual exposures of $11.7 - 91.9$ kiloseconds.  Three of the clusters in our sample were observed on multiple occasions, in which case we have chosen the longest of those observations to include in our analysis.  Each of the observations analyzed in this study possesses a start date which falls on or after 24 April, 2000, indicating a focal plane temperature of $-120^o$ C for the entire sample.  

Aspect solutions were examined for irregularities and none were
found.  Background contamination due to charged particle flares were
reduced by removing time intervals during which the background rate
exceeded the average background rate by more than $20\%$.  The event files were filtered on
standard grades and bad pixels were removed.  A two-dimensional elliptical Lorentzian was fit to the counts image of each dataset to locate the center of the X-ray emission peak.  All centroid position errors are within a resolution element ($\sim0.5\arcsec$).  In the case of the heavily substructured cluster MS0451+02, a fitted central peak was unobtainable, so a position at the center of the extended emission was chosen for spectral analysis at an RA, Dec of 04:54:09.941,+02:55:14.52 (J2000), and the surface brightness profile was centered on the slightly offset peak of emission at an RA, Dec of 04:54:07.249, +02:54:27.31 (J2000).

Table~\ref{table1} provides a list of each of the clusters in our sample, including redshifts, obs-id, detector array, and corrected exposure information for each observation. 

After initial cleaning of the data, 0.6-7.0 keV images, instrument maps, and exposure maps were created using the CIAO 3.2 tools DMCOPY, MKINSTMAP and MKEXPMAP.  Data with energies below 0.6 keV and above 7.0 keV were excluded due to uncertainties in the ACIS calibration and background contamination, respectively.

Flux images were created by dividing the resulting images by their respective exposure maps.  Point source detection was performed by running the tools WTRANSFORM and WRECON on the flux images.  Adaptively smoothed flux images were created with CSMOOTH.  Figure~\ref{fig1} contains CFHT images obtained from the CNOC database \citep{yee96}, and, where available, HST images obtained from the MAST archive, each of which is overlayed with X-ray contours created from smoothed Chandra flux images.

\section{Surface Brightness\label{s:sb}}

Using the 0.6-7.0 keV images and exposure maps, a radial surface brightness profile was computed in $1\arcsec$ annular bins for each cluster.  These profiles
were then fit with both single and double $\beta$ models.  Single
$\beta$ models take the form 

\begin{equation}
I(r) = I_B + I_0 \left( 1 + {r^2 \over r_c^2} \right)^{-3\beta+\frac{1}{2}},
\label{sb_eq}
\end{equation} 
where $I_B$ is a constant representing the surface brightness contribution of the background, 
$I_{0}$ is the normalization and $\rm{r}_{\rm{c}}$ is the core
radius.

The double $\beta$-model has the form
\begin{equation}
I(r) = I_B +
       I_{1} \left( 1 + {r^2 \over r_{1}^2} \right)^{-3\beta_1+\frac{1}{2}} + 
       I_{2} \left( 1 + {r^2 \over r_{2}^2} \right)^{-3\beta_2+\frac{1}{2}},
\label{sbdoub_eq}
\end{equation}
where each component has fit parameters $(I_n, r_n, \beta_n)$.

Though a fair number of the clusters in our sample posess detectably elliptical or irregular emission, circular surface brightness profiles were chosen based on a number of factors.  First, as discussed in~\citet{neumann} and~\citet{bohringer}, differences in masses derived using elliptical vs. circular $\beta$ profiles are small ($\sim5\%$).  Secondly, circular $\beta$ profiles provide a more straightforward comparison between dynamical and X-ray derived masses, since dynamical mass calculations assume sphericity.  Finally, temperature uncertainties should outweigh any errors introduced by assuming radial symmetry.

The parameters of the best fitting single $\beta$ model of each cluster are shown in Table \ref{table2}.  While nine of the clusters' surface brightness profiles do not exhibit excess unmodeled emission in their cores when fit with a single $\beta$ model, five of the clusters do exhibit excess emission in their cores.  The surface
brightness profiles of these clusters are better fit with the addition
of a second $\beta$ component.  All of the clusters requiring a double
$\beta$ model are believed to contain cooling flows: Abell 2390, MS0440+02, MS0839+29, MS1358+62, and MS1455+22~\citep{lewis,allen00} and exhibit strong ($L_{\rm{H}\alpha+[\rm{N}_{\rm{II}}]}>10^{42}~\rm{erg}~\rm{s}^{-1}$) extended H$\alpha$ emission~\citep{donahue92,lewis}.  Best fitting double $\beta$ model parameters are given in Table~\ref{table3}.  Surface brightness profiles of the clusters in our sample are shown in Figure~\ref{fig2}.  Discrepancies in background values are due to the use of both ACIS-I and ACIS-S arrays (ACIS-S typically has background values which are a factor of three higher than ACIS-I).

\section{Spectral Analysis of X-ray Observations\label{s:spect}}

\subsection{Integrated Spectra and {$\rm{R}_{2500}$}\label{ss:integrated}}

For the purpose of fitting cluster scaling relationships within a well defined region of mass overdensity, global cluster properties were determined for our sample within $\rm{R}_{2500}$.  Using the fitted X-ray centroid positions obtained in Section~\ref{s:obs} (with the exception of MS0451+02, as noted in Section~\ref{s:obs}), a
spectrum was extracted from each cleaned event file in a circular region with
a $300~\rm{h}_{70}^{-1}~\rm{kpc}$ radius.  These spectra were then analyzed with XSPEC 11.3.2~\citep{arnaud96}, using weighted response matrices (RMFs) and effective area files (ARFs) generated with the CIAO tool ACISSPEC and the latest version of the Chandra calibration database (CALDB 3.0.1). Background spectra were extracted from the aimpoint chip as far away from the cluster as possible. Each
spectrum was grouped to contain at least 20 counts per energy bin.  

Spectra were fitted with single temperature spectral models, inclusive of
foreground absorption.  Each spectrum was initially fit with the absorbing column frozen at its measured value~\citep{dickey90}, and redshifts were fixed throughout the analysis.  Metal abundances were allowed to vary.  Data with energies below 0.6 keV and above 7.0 keV were again excluded.  

In some cases evidence of excess photoelectric absorption has been seen in cooling flow clusters~\citep{allen97,allen00}.  To investigate this possibility a second fit was performed, allowing the absorbing column to vary.  No evidence for excess absorption was found in 13 of the 14 clusters.  The MS0440+02 spectrum, however, was equally well fit by a model inclusive of excess foreground absorption.  

The results of these fits, combined with the best fitting $\beta$ model parameters from Section~\ref{s:sb}, were then used to estimate the value of $\rm{R}_{2500}$ for each cluster.  This is accomplished by combining the equation for total gravitating mass

\begin{equation}
M_{tot}(<r) = -{{T(r)r}\over{G \mu m_p}} \left({{\delta~\rm{ln}~\rho}\over{\delta~\rm{ln}~r}} + {{\delta~\rm{ln}~T} \over {\delta~\rm{ln}~r}} \right), 
\label{Mass_eq}
\end{equation}
 
\noindent
(where $\mu m_p$ is the mean mass per particle) with the definition of mass overdensity

\begin{equation}
M_{tot}(r_\Delta) = {{4}\over{3}}\pi\rho_c r^3_\Delta \Delta ,
\end{equation}

\noindent
where the critical density $\rho_c = {{3 H_z^2}/{8 \pi G}}$, $z$ is the cluster redshift, and $\Delta$ is the factor by which the density at $r_\Delta$ exceeds the critical density.  These equations are then combined with the density profile implied from the $\beta$ model (assuming hydrostatic equilibrium, spherical symmetry, and isothermality)

\begin{equation}
\rho_{gas}(r) = \rho_0 \left[1 + {{r^2}\over{r_c^2}}\right]^{-3\beta/2},
\label{dens_eq}
\end{equation}

\noindent
resulting in the equation

\begin{equation}
{{r_\Delta}\over{r_c}} = \sqrt{{\left[{{3\beta T}\over{G \mu m_p (4/3) \pi \rho_c (1+z)^3 r_c^2 \Delta}}\right]}-1},
\label{eq:ettori}
\end{equation}

\noindent
\citep{ettori00,ettori04}.

After the initial estimation of $\rm{R}_{2500}$, additional spectra were extracted from within that radius, and spectral fitting was performed again.  This iterative process was continued until fitted temperatures and calculated values of $\rm{R}_{2500}$ were consistent with extraction radii.  The results of this process, including values of $\rm{R}_{2500}$, temperatures and abundances are shown in Table~\ref{table4}, along with 90\% confidence ranges.

An additional process was carried out for the five cooling core clusters in our sample.  Using spatially resolved temperature profiles (Section~\ref{ss:annuli}), the radius within which cooling became prominent in each cluster ($\rm{R}_{\rm{cutoff}}$) was estimated.  A single annular spectrum was then extracted for each cluster from that radius to $\rm{R}_{2500}$ and fitted with single temperature spectral models.  The observation of MS0440+02 did not possess enough signal to constrain a fit to the cooling flow excised spectrum, therefore it was discarded.  The results of these fits are shown in Table~\ref{table5}, along with inner extraction radii and calculated values of $\rm{R}_{2500}$.   

Overall, our integrated temperatures are comparable to the previous ROSAT results of~\citet{lewis}, though they are better constrained through Chandra observations.  More recent analyses of subsets of these clusters have been conducted with ROSAT~\citep{mohr}, ASCA~\citep{matsumoto,henry04}, and Chandra ~\citep{allen01,arabadjis,donahue03,ettori03,ettori04}.  Our fitted temperatures are consistent within errors of the vast majority of the values reported in these papers.  Noticeable discrepancies arise occasionally in the case of cooling flow clusters (where some of these authors have not excised the cooling flow contribution to integrated temperature) and MS1008-1224, for which we consistently derive a lower temperature than other studies.  This is not surprising considering the irregular morphology of this cluster, and may be due to differences in the regions for which spectra were extracted.  Metallicities for this sample are consistent with an overall value of $\sim 0.3$ solar, as is seen in lower redshift clusters.  This suggests that ICM enrichment occurs early in a cluster's history, evidence which is corroborated by the work of~\citet{richard}.

\subsection{Spatially Resolved Spectra\label{ss:annuli}}

With the goal of elucidating the radial dependence of temperature for the clusters in our sample, spectra were extracted within circular annuli spanning the central $150-600~\rm{h}_{70}^{-1}~\rm{kpc}$ of each cluster, depending on the signal-to-noise ratio of each observation.  The extraction regions were sized to include at least 2500 counts per spectrum in the 0.6-7.0 keV band for each data set. This number was chosen as a minimum for achieving reasonable temperature constraints on the emission in each annulus. 

Extractions were performed after the removal of previously detected point sources (Section~\ref{s:obs}), using the CIAO tool ACISSPEC, which incorporates the changing response over large areas of the detector.  Three of the 14 clusters in this sample did not possess enough signal for more than one extraction region, and were removed from the sample during the subsequent spatial analysis.  Spectra from the eleven remaining clusters were then grouped to contain 20 counts per bin, and were analyzed using XSPEC.  Single temperature spectral models with fixed galactic absorption and varying abundances were fitted to the data.  The results of these fits were used to create a temperature profile for each cluster.  

Figure~\ref{fig3} illustrates the eleven resulting temperature profiles.  It is clear from these figures that significant cooling cores are present in five of the clusters in this sample: Abell 2390, MS0440+02, MS0839+29, MS1358+62, and MS1455+22.  These are the same five clusters which required double $\beta$ model fits to their surface brightness profiles (Section~\ref{s:sb}).  The remaining six temperature profiles are consistent with isothermality.

\subsection{The Universal Temperature Profile}

The spatially resolved temperature profiles of the five cooling core clusters were scaled by $\rm{R}_{2500}$ and $\rm{T}_{2500}$ in an attempt to check their consistency with the universal temperature profile proposed by~\citet{allen01}.  The combined data from these five clusters were then fit using the functional form $\rm{T(R)}/\rm{T}_{2500} = \rm{T}_0 + \rm{T}_1[(x/x_c)^\eta/(1+(x/x_c)^\eta)]$, with $x=\rm{R}/\rm{R}_{2500}$~\citep{allen01}.  This fit resulted in $\rm{T}_0=0.397\pm{0.007}$, $\rm{T}_1=0.56\pm{0.02}$, $x_c=0.0865\pm{0.0002}$, and $\eta=1.3\pm{0.2}$, with a reduced $\chi^2$ of 1.3.

These values are all consistent with the best fit found by~\citet{allen01}.  The resulting function, however, does not asymptotically approach 1 at large radii, ostensibly due to the fact that these temperature profiles do not extend all the way to $\rm{R}_{2500}$.  Figure~\ref{fig4} shows the individual temperature profiles as well as the binned and averaged temperature profile of the five clusters, with the best fitting functions overlayed.

\section{Mass Determinations\label{s:mass}}

Using the results of the spectral fits and $\beta$ model fits, along with Equations~\ref{Mass_eq},~\ref{dens_eq}, and~\ref{eq:ettori}, gas masses and total masses were calculated out to both $\rm{R}_{2500}$ and $R_{200}$ for the clusters in our sample.  Central densities were determined via an expression relating the observable cluster X-ray luminosity to gas density, using emission measures obtained during spectral fitting in XSPEC.

For the nine non-cooling core clusters, single temperature spectra fits and single $\beta$ model parameters were used.  For the five clusters which exhibit significant cooling, spectral parameters from cooling core excised spectral fits were used, and $\beta$ and $r_{c}$ were taken from the results of double $\beta$ model fits.  

To determine the effects of inner temperature gradients on total cluster mass, least-squares fitting was performed on the temperature profiles obtained in Section~\ref{ss:annuli} for the five cooling core clusters.  The resulting parameterizations were included in Equation~\ref{Mass_eq}, and masses were calculated out to the edge of the cooling region (Table~\ref{table5}).  Masses were also calculated without the inclusion of this parameterization, and the results compared.  According to the outcome of this exercise, the inclusion of temperature gradients in mass calculations of the five cooling core clusters in our sample would result in, at most, a $0.3\%$ correction to the total mass within the cooling region.  This correction is negligible compared to the other uncertainties in mass calculations and is therefore not included in the final results.


An additional uncertainty is present in mass estimations of MS0451+0250 due to its significant irregularity.  Choosing a centroid at the center of the extended emission rather than one at the most central peak of emission produces a total mass which is greater by 25\%.  The beta parameter which results from using this centroid, however, is unusually high (1.9).  X-ray mass determinations are presented in Tables~\ref{table6a} and~\ref{table6} along with $68\%$ confidence intervals.

\section{Gas Mass Fractions and $\Omega_m$\label{s:gasmass}}

Under the assumption that clusters provide a fair representation of the universe, gas mass fractions, $f_{gas}$, defined as the ratio of cluster gas mass to total gravitating mass, were calculated for our sample within $\rm{R}_{200}$ and are listed in Table~\ref{table6}.  $f_{gas}$ can be used to calculate the cosmological mass density, $\Omega_m$, via the relation

\begin{equation}
\Omega_m = {{\Omega_b}\over{f_{gas}~(1+0.19h^{0.5})}},
\label{omegam}
\end{equation}

\noindent
\citep{allen02}, where $0.19h^{0.5}$ represents the baryonic contribution from optically luminous matter~\citep{white93,fukugita}.  A value of $\Omega_b~h^{2}$ of $0.0223^{+0.0007}_{-0.0009}$ is adopted~\citep[WMAP,][]{spergel2}, and here we take $h=0.71$.  Using an average $f_{gas} = 0.136\pm{0.004} h_{70}^{-3/2}$ within $\rm{R}_{200}$ for our sample, we calculate a cosmological mass density of $\Omega_m = 0.28\pm{0.01}$ (68\% confidence, with error bars representing the formal error).  This value is in good agreement with WMAP three year results~\citep{spergel2}.

\section{Mass Comparisons\label{s:masscomp}} 

\subsection{Dynamical Mass Comparisons}

Comparisons between X-ray and dynamical masses for the CNOC sample were previously undertaken by~\citet{lewis} using ROSAT observations.  While they found good agreement between the two methods of mass estimation, accurate surface brightness modeling and detailed investigations of cluster temperature gradients were unavailable due to the comparitively poor spatial resolution of ROSAT (particularly at moderate redshift).  Here we utilize the $0.5\arcsec$ spatital resolution of Chandra to improve the accuracy of these initial comparisons.

Detailed dynamical studies of CNOC clusters were performed by~\citet{carlberg},~\citet{borgani}, and~\citet{vandermarel}. ~\citet{carlberg} provides velocity dispersions obtained from CFHT spectroscopy,~\citet{borgani} adjusts these values by employing an improved interloper-removal algorithm, and~\citet{vandermarel} investigates the isotropicity and galaxy distribution of a composite CNOC cluster created via dimensionless scaling.  Here we will primarily draw from the work of~\citet{borgani} and~\citet{vandermarel} for our mass estimates.

Dynamical masses can be calculated from velocity dispersions via the Jeans equation

\begin{equation}
M(r) = {{-\sigma^2_r r}\over{G}}\left({{\delta~\rm{ln}~\sigma_r^2} \over {\delta~\rm{ln}~r}} + {{\delta~\rm{ln}~\nu(r)} \over {\delta~\rm{ln}~r}} + 2\beta \right),
\label{virial}
\end{equation}

\noindent 
where $\sigma_r$ represents the radial velocity dispersion, $\nu(r)$ is the galaxy number density profile, and $\beta$ represents the anisotropy of the system.  According to~\citet{vandermarel}, the CNOC clusters can be treated as isotropic (i.e. $\beta = 0$), and $\nu(r)$ takes the form

\begin{equation}
\nu(r)=\nu_0 (r/a)^{-\gamma} \left[1+r/a\right]^{\gamma-3},
\end{equation}

\noindent
where the length scale of the mass distribution is set by the parameter $a$, and $\gamma$ represents the logarithmic power-law slope near the center.  The best fitting values of $a$ and $\gamma$ for isotropicity are 0.224 and 0.75, respectively.  This relationship was obtained by creating a composite CNOC cluster via dimensionless scaling~\citep{vandermarel}, as was a plot of $\sigma$ vs. $r$.  This figure~\citep[Figure 2]{vandermarel} indicates that $\sigma$ is not a strong function of radius out to $R_{200}$, and in keeping with both~\citet{lewis} and~\citet{carlberg} we assume that $\sigma(r)=\sigma$.

Using Equation~\ref{virial}, dynamical masses were calculated for our sample out to $R_{200}$ (as determined by X-ray parameters in Section \ref{ss:integrated}), using the velocity dispersions of~\citet{borgani}.  These masses were then compared to X-ray derived masses from the previous section (Section~\ref{s:mass}).  A weighted average gives an overall dynamical to X-ray mass ratio of $0.97\pm0.05$, where the error bar indicates the uncertainty in the mean.  Table~\ref{table7} lists both dynamical and X-ray derived masses along with their ratio and $68\%$ confidence intervals.  Figure~\ref{fig5} is a plot indicating the dynamical to X-ray mass ratio of each cluster in the sample. 

The high dynamical to X-ray mass ratios of MS1006.0+1202 and MS1008.1-1224 may be due to overestimated velocity dispersions of these objects, as they both have markedly irregular emission (Figure~\ref{fig1}).  However we also see evidence of good agreement between X-ray and dynamical masses in irregular objects (MS0451.5+0250), as well as disagreement in some regular objects (MS0839.8+2938), therefore a clear pattern does not make itself evident.  Likewise, cooling core objects show no obvious systematic departures from consistency.  Overall, dynamical and X-ray mass estimations for this sample show remarkable agreement.

Our resulting ratio of dynamical to X-ray masses is consistent with that quoted by~\citet{lewis} in their ROSAT study, however the scatter about the mean of our distribution is $\sim 5\%$ smaller.  This decrease in scatter is indicative of the improved spatial and spectral resolution of Chandra.  In addition,~\citet{lewis} systematically overestimate the core radii of cooling flow clusters, and therefore their masses, another result which is likely due to the poorer spatial resolution of ROSAT. 

\subsection{Weak Lensing Mass Comparisons}

Weak lensing mass estimates were obtained for seven of the clusters in our sample using deep optical observations at the CFHT 3.6m telescope~\citep[in preparation]{hoekstra}.  The model-independent projected mass estimates that we employ in this paper were calculated for the inner $500~h^{-1}~\rm{kpc}$ of each cluster, using a cosmology of $H_0=100$, $\Omega_m=0.3$ and $\Omega_\Lambda=0.7$.

To compare X-ray derived masses to weak lensing masses, we calculated a cylindrical X-ray mass within $500~h^{-1}~kpc$.  This was done using previously determined $\beta$ models, the adopted cosmology (above), and a cylindrical mass projection out to $10~\rm{Mpc}$ from the cluster core.  A comparison of weak lensing masses to X-ray derived masses can be found in Table~\ref{table8}, a plot of mass ratios is given in Figure~\ref{fig6}.  Contributions to lensing signal from structures along the line of sight may result in masses which are biased somewhat high.  Similarly, large scale structure along the line of sight results in increased scatter~\citep{hoekstra01}.  Despite these possible challenges, the weighted average of our mass ratios gives a weak lensing to X-ray mass ratio of $0.99\pm0.08$, with a reduced $\chi^2$ of 0.93.  Though the distribution in Figure~\ref{fig6} appears assymmetric, it is not statistically significant.

\section{X-ray Scaling Laws\label{s:laws}}

\subsection{The $L_X-T_X$ Relationship\label{ss:lxtx}}

Unabsorbed bolometric X-ray luminosities within $\rm{R}_{2500}$ were obtained for our sample in the following manner.  Unabsorbed X-ray luminosities for the 2-10 keV band were calculated in XSPEC during spectral fitting (Section~\ref{ss:integrated}).  In the case of the five clusters which exhibit significant cooling, corrected, integrated luminosities were obtained by fixing X-ray temperatures at the values determined for cooling core corrected spectra (Table~\ref{table5}).  To convert 2-10 keV luminosities to bolometric X-ray luminosities, correction factors were obtained via NASA's
Portable, Interactive Multi-Mission Simulator (PIMMS) for each individual cluster, using a thermal bremsstrahlung model and the spectrally determined temperature of the cluster.  The resulting bolometric luminosities are given in Table~\ref{table9}.

The $L_X-T_X$ data were fit using the bivariate correlated errors and intrinsic scatter (BCES) estimator of~\citet{akritas}.  This estimator allows for measurement errors in both variables as well as possible intrinsic scatter, and was used in both~\citet{allen01} and~\citet{yee03}.  To correct for cosmological effects, a form similar to~\citet{allen01} was adopted:

\begin{equation}
\left({L_{2500}}\over{10^{44}~\rm{erg}~\rm{s}^{-1}~E(z)}\right)=C_1 \left(\rm{kT}_{2500}\right)^{C_2},
\end{equation}

\noindent where $E(z) = {H_z}/{H_0} = \left[{\Omega_m (1+z)^3 + \Omega_\Lambda} \right]^{1/2}$.  The best fitting values were $C_1 = 0.13^{+0.08}_{-0.05}$ and $C_2 = 2.4\pm{0.2}$.  While this slope is lower than some estimates~\citep[e.g.][]{arnaud99,ettori04}, it is consistent with both~\citet{allen01} and~\citet{ettori02}.  Because it is steeper than the expected self-similar slope of 2~\citep{voit}, it indicates modest negative evolution in the $L_X-T_X$ relationship at $z\sim0.3$.

\citet{ettori04} examine scaling laws in clusters at moderate to high redshift ($0.4<z<1.3$), so here we adopt their definition of scatter for comparison 

\noindent($\sigma_{Y} \equiv \left[\sum_{j=1,N}(log~L_{\rm{j,obs}}-log~L_{\rm{j,pred}})^2/N\right]^{1/2}$).  Defined this way, our scatter in luminosity $\sigma_{Y}=0.20$, is smaller than that of Ettori et al. ($\sigma_Y=0.35$); however they have included significantly higher redshift clusters in their fit.  A plot of the data with our relationship overlaid is presented in Figure~\ref{fig7}.

\subsection{The Mass$-T_X$ Relationship\label{ss:masstx}}

A BCES fit was also performed between cluster temperatures and mass estimates.  This relationship takes the form \citep{ettori04}:

\begin{equation}
\left({{E(z)}{M_{tot}}}\right)=C_1 \left(\rm{kT}\right)^{C_2}.
\end{equation}

\noindent 
We again use cluster properties which were determined within $\rm{R}_{2500}$.  The best fitting values were $C_1 = 1.3^{+0.3}_{-0.2} \times 10^{13}$ and $C_2 = 1.7\pm{0.1}$.  This slope is consistent with~\citet{allen01},~\citet{ettori02}, and~\citet{ettori04}, and is again steeper than the expected slope of 1.5~\citep{voit}.  Our scatter in mass is also lower than~\citet{ettori04} ($\sigma_{Y}=0.07$ compared to $\sigma_Y=0.15$).  A plot of the data with our relationship overlaid is presented in Figure~\ref{fig8}.

\section{X-ray Properties and Optical Richness\label{s:or}}


Optical richness is in essence a measurement of the galaxy excess in the direction of a cluster above a certain magnitude limit and within a specific aperture.  The particular optical richness parameter that is used in this work, $B_{gc}$, is defined as the amplitude of the galaxy-cluster correlation function ~\citep{long}:

\begin{equation}
\xi (r) = \left({{r}\over{r_0}}\right)^{-\gamma} = ~{\rm{B}}_{gc}~r^{-\gamma}
\end{equation}

Galaxies are observed as projections on the sky, therefore what is measured is the angular two-point correlation function of galaxies, $\omega(\theta)$, where $\theta$ is the angle on the sky.  $\omega(\theta)$ is approximated by a powerlaw of the form: 

\begin{equation}
\omega(\theta)=A_{\rm{gg}}\theta^{1-\gamma}
\end{equation}

\noindent\citep{davis,yee99}, where $A_{\rm{gg}}$ is the galaxy-galaxy angular correlation amplitude.  $\omega_\theta$ is taken as the the distribution of galaxies around the the center of the cluster, and the amplitude is then relabeled as $A_{\rm{gc}}$.  This amplitude can be measured from an image by counting the background-corrected excess of galaxies within a certain $\theta$ to a particular magnitude limit.

$\rm{B}_{gc}$ is then calculated through a deprojection analysis which assumes spherical symmetry~\citep{long}:

\begin{equation}
\rm{B}_{gc}=N_{bg} {{D^{\gamma-3} A_{\rm{gc}}}\over{I_\gamma \Phi[M(m_0,z)]}}
\end{equation}

\noindent where $N_{bg}$ represents the background galaxy counts to apparent magnitude $m_0$, D is the angular diameter distance to the cluster redshift $z$, $I_\gamma$ is an integration constant, and $\Phi[M(m_0,z)]$ is the integrated luminosity function of galaxies to the absolute magnitude $M$ which corresponds to $m_0$ at $z$.

One of the challenges in the calculation of $B_{gc}$ involves the lack of a complete knowledge of the galaxy luminosity function at high redshifts.  This uncertainty can be minimized by employing the parameter $B_{gc,red}$~\citep{gladders05}, which is calculated using galaxies in the red-sequence.  This parameter is expected to provide a more robust indication of cluster mass due to the well understood passive evolution of red-sequence cluster galaxies~\citep{vandokkum} as opposed to the more unpredictable nature of star forming populations.  Since one of our main goals in this work is to provide a comparison sample for future studies of clusters at higher redshift, we use the parameter $B_{gc,red}$ for the measurement of richness.  

At the relatively low redshifts of the CNOC clusters, the difference between $B_{gc}$ and $B_{gc,red}$ is small.  We estimate $B_{gc,red}$ values by applying small corrections to the original $B_{gc}$ values of the CNOC clusters~\citep{yee03}, based on their blue fractions~\citep{ellingson}.  Most of the corrections are of the order of $\sim 10\%$.  Values of $B_{gc,red}$ (in units of $\rm{h}_{50}^{-1}~\rm{Mpc}^{1.77}$), as well as the cluster X-ray luminosities obtained in Section~\ref{ss:integrated}, are given in Table~\ref{table9}.  To keep the same scale as previous work using the $B_{gc}$ parameter, they are computed using $\rm{H}_0=50~\rm{km}~\rm{s}^{-1}~\rm{Mpc}^{-1}$.

$B_{gc}$ has previously been shown to correlate strongly with the X-ray properties of clusters in this sample~\citep{yee03}.  Here we re-examine these correlations using improved X-ray data from Chandra.  Extending the simple relationships expressed in~\citet{yee03} to $B_{gc,red}$ we have

\begin{equation}
B_{gc,red} \propto T_{X}^{\gamma/2},
\label{eq:txbgc}
\end{equation}

\noindent and using our best fit $L_{X}-T_{X}$ and Mass$-T_X$ relationships (Section~\ref{s:laws}), in combination with Equation~\ref{eq:txbgc}

\begin{equation}
B_{gc,red} \propto {\left({L_{X}}\over{E(z)}\right)}^{\gamma/4.8},
\label{eq:lbgc}
\end{equation}

\noindent and

\begin{equation}
B_{gc,red} \propto {\left(E(z)~\rm{M}_{2500}\right)}^{\gamma/3.4},
\label{eq:mbgc}
\end{equation}

\noindent In this section we derive the best fitting relationships between cluster X-ray properties and $B_{gc,red}$ for our sample.  A generic form for these relationships of

\begin{equation}
\rm{log~X} = C_1 + C_2~\rm{log}~{B_{\rm{gc,red}}},
\label{eq:powerlaw}
\end{equation}

\noindent is adopted, where X represents the particular property being fit.  For $T_X$, $E(z)~\rm{M}_{2500}$, and $L_X/E(z)$, units of keV, $\msun$, and $10^{44} \rm{erg}~\rm{s}^{-1}$ were used, respectively.  As in~\citet{yee03}, the BCES estimator of~\citet{akritas} was employed.

The results of these fits are all consistent within errors with those of~\citet{yee03}, and are also consistent with the expected value of $\gamma$ ($\gamma=1.77$).  The data, with the results of these fits as well as those of~\citet{yee03} overlaid, are shown in Figures~\ref{fig9}-\ref{fig11}.  Best fitting parameters of all fits are given in Table~\ref{table10}.

\section{Summary and Discussion\label{sum}}

We have presented a comprehensive analysis of Chandra observations of 14 medium redshift ($0.1709<z<0.5466$) clusters of galaxies from the CNOC subsample of the EMSS (Table~\ref{table1}). Imaging analysis has provided information on the relative quiescence of each cluster (Figure~\ref{fig1}).  The spatial resolution of Chandra has allowed us to determine the surface brightness profiles of these clusters down to $1\arcsec$ scales (Figure~\ref{fig2}).  This has enabled us to obtain precise models of cluster emission (Table~\ref{table2}), except in the case of MS0451+02 which displays a considerable amount of substructure.  Five clusters in our sample (Abell 2390, MS0440+02, MS0839+29, MS1358+62, and MS1455+22) exhibit excess emission in their cores, indicating the presence of cool gas.  The surface brightnesses of these five clusters are better fit with the inclusion of a second $\beta$ component to model the excess core emission (Table~\ref{table3}).

Chandra's $0.5\arcsec$ spatial resolution has enabled us to study the radial distribution of temperatures on small scales ($\sim7-100$ kpc) in the inner 150-600 kpc of eleven of these objects (Figure~\ref{fig3}).  While nine clusters in our sample have temperature profiles which are consistent with isothermality, the five cooling core clusters (listed above) show clear temperature gradients in their innermost $\sim60-200$ kpc.  The temperature profiles of these five clusters are consistent with the ``universal'' temperature profile of \citet{allen01}.

The energy resolution of the instrument has provided well constrained spectral analyses of the central $\rm{R}_{2500}$ ($\sim350-760$) kpc of each object.  Temperatures obtained initially from 300 kpc radius extraction regions began an iterative spectral extraction and fitting process culminating in the determination of robust integrated temperatures, luminosities, and abundances within $\rm{R}_{2500}$ (Tables~\ref{table4} and~\ref{table5}).  Cooling core corrected spectra were used to determine global temperatures for the clusters in our sample which exhibit significant cooling.  These temperatures were then employed in the spectral determination of integrated luminosities for these objects.  Overall, our sample displays temperatures in the range of $3.4-12.1$ keV, abundances which range from $0.1-0.8$ times the solar value, and unabsorbed bolometric luminosities within $\rm{R}_{2500}$ which span $4.9-56.1 \times 10^{44}~\rm{erg}~\rm{s}^{-1}$.

Cluster gas and total masses within $\rm{R}_{2500}$ and $\rm{R}_{200}$ were estimated using the outcomes of spectral and surface brightness fitting, resulting in virialized cluster masses of $2.7-21 \times 10^{14}~h_{70}^{-1}\msun$ (Table~\ref{table6}) and respective cluster gas masses of $0.13-2.9 \times 10^{14}~h_{70}^{-1}\msun$.  A weighted average of gas mass fractions gives $f_{gas}=0.136\pm{0.004}~h_{70}^{-3/2}$, resulting in $\Omega_m=0.28\pm{0.01}$ (68\% confidence, formal error), this value is in good agreement with~\citet{spergel2}.  

Dynamical masses within $\rm{R}_{200}$ were calculated via the Jeans Equation using velocity dispersions taken from~\citet{borgani} (Table~\ref{table7}).  Comparisons between X-ray masses and dynamical masses result in a weighted average of $\rm{M}_{\rm{dyn}}/\rm{M}_{\rm{X-ray}}=0.97\pm{0.05}$, indicating good agreement between between these two methods, however a fair amount of scatter is evident ($\sim 30\%$; Figure~\ref{fig5}).  Dynamical masses are noticeably larger in the case of clusters which exhibit significant substructure (MS1006+12 and MS1008-12), a factor which may be responsible for an overestimation of velocity dispersions~\citep{bird}.

Weak lensing masses within $500~h_{100}^{-1}$ kpc were obtained for seven of the clusters in our sample.  X-ray masses were calculated within this region and compared to the lensing results (Table~\ref{table8}).  Although the distribution appears somehwat asymmetric (Figure~\ref{fig6}), a weighted average gives $\rm{M}_{\rm{lens}}/\rm{M}_{\rm{X-ray}}=0.99\pm{0.07}$, with a scatter of $\sim 30\%$.  

X-ray scaling laws for this sample were investigated in a manner similar to~\citet{allen01} and~\citet{ettori04}, taking cosmological factors into account.  The best fitting $L_X-T_X$ relationship for our sample results in an intercept of $0.13^{+0.08}_{-0.05}$ and a slope of $2.4\pm{0.2}$ (Figure~\ref{fig7}).  While this slope is lower than some estimates~\citep{arnaud99,ettori04}, it is consistent with both~\citet{allen01} and~\citet{ettori02}.  Moderate negative evolution is indicated by this index being greater than that predicted by self-similar models~\citep{voit}.  The Mass-$T_X$ relationship for these clusters also exhibits a somewhat steeper slope than expected, at $1.7\pm{0.1}$, with an intercept of $1.3^{+0.3}_{-0.2} \times 10^{13}$ (Figure~\ref{fig8}).

The best fitting scaling laws for our sample (listed above) were combined with Equation~\ref{eq:txbgc}, with the ultimate goal of calibrating relationships between red-sequence optical richness ($\rm{B}_{gc,red}$) and global cluster parameters ($T_X$, $L_X$ and $\rm{M}_{2500}$).  We find that $T_X$ scales relatively well to $\rm{B}_{gc,red}$ with a $40\%$ scatter (Figure~\ref{fig9}).  $\rm{M}_{2500}$ shows a scatter of $\sim70\%$ (Figure~\ref{fig11}), which is consistent with the $T_X$-$\rm{B}_{gc,red}$ scatter given that ${\rm{M}}\propto T^{3/2}$.  The $L_X$-$\rm{B}_{gc,red}$ relationship exhibits a significantly larger scatter at $\sim 200\%$ (Figure~\ref{fig10}).  

Our results indicate that $\rm{B}_{gc,red}$ does exhibit intial promise as a mass indicator.  Accurate calibration of a relationship between optical richness and cluster mass will require the use of additional clusters which posess both well-constrained X-ray temperatures and optical richness measurements.

Overall we find that multiple cluster mass estimators - dynamics, weak lensing and X-ray observations, along with optical richness in the cluster red sequence - are converging for this sample of well-studied clusters.  While individual correlations still have significant scatter, there is little evidence for large systematic bias in any of these methods.  Cluster characteristics which might be considered problematic for one or more techniques (cluster substructure, merging, and/or the presence of a cool core) appear to perturb these relationships relatively little, as long as high quality data are obtained and the analysis is tuned to correct for these.  In particular the correlation between cluster temperature and optical richness, the most easily obtained of the mass estimators, is promising.  A remaining concern is that X-ray selected clusters may not prove to be typical of all massive clusters at these redshifts.  Additional checks involving clusters covering a broader range of redshift and selection technique (e.g., SZ, optical, weak lensing) will be necessary to solidify our understanding of the most reliable and efficient methods of cluster mass estimation.

\acknowledgements 

Support for this work was provided by the National Aeronautics and Space Administration through a Graduate Student Research Program (GSRP) fellowship, NGT5-140, and Chandra Awards GO0-1079X and GO0-1063B, issued by the Chandra X-ray Observatory Center, which is operated by the Smithsonian Astrophysical Observatory for and on behalf of the National Aeronautics Space Administration under contract NAS8-03060.  We would also like to thank Phil Armitage, Mark Bautz, Webster Cash, John Houck, and Richard Mushotzky for their contributions and input.

Some of the data presented in this paper were obtained from the Multimission Archive at the Space Telescope Science Institute (MAST). STScI is operated by the Association of Universities for Research in Astronomy, Inc., under NASA contract NAS5-26555. Support for MAST for non-HST data is provided by the NASA Office of Space Science via grant NAG5-7584 and by other grants and contracts.

\clearpage



\begin{figure}
\includegraphics{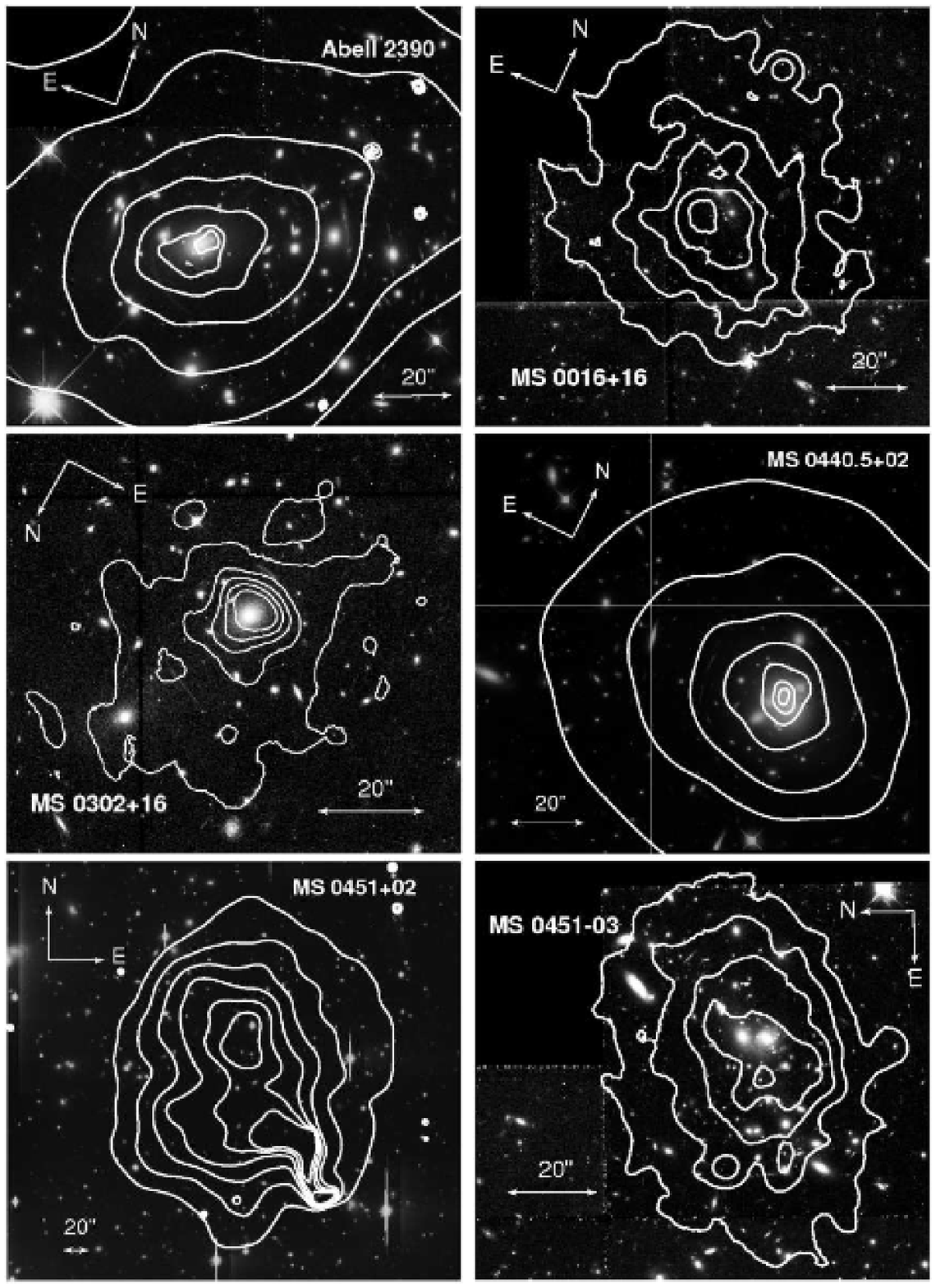}
\end{figure}
\begin{figure}
\includegraphics{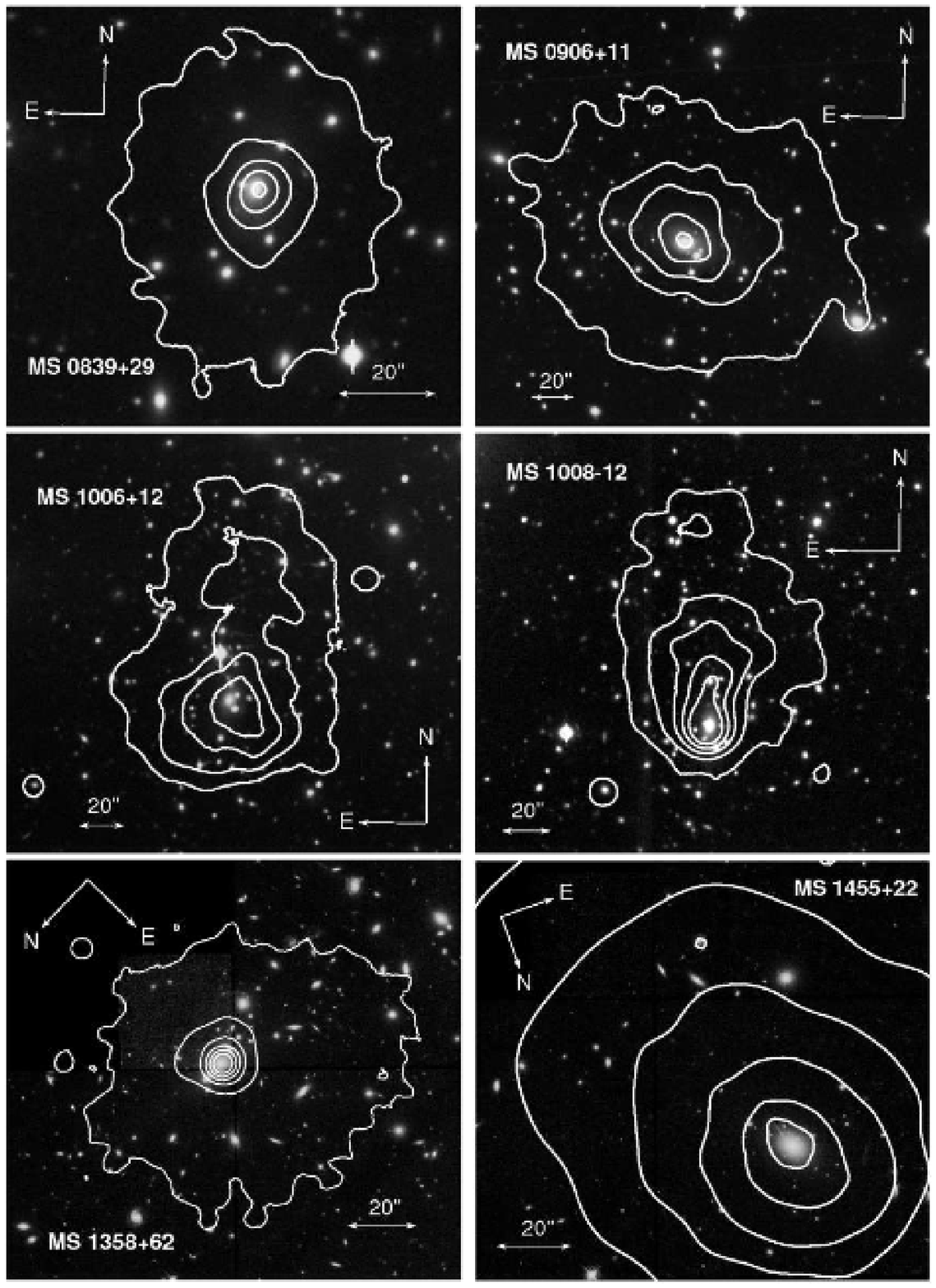}
\end{figure}
\begin{figure}
\includegraphics{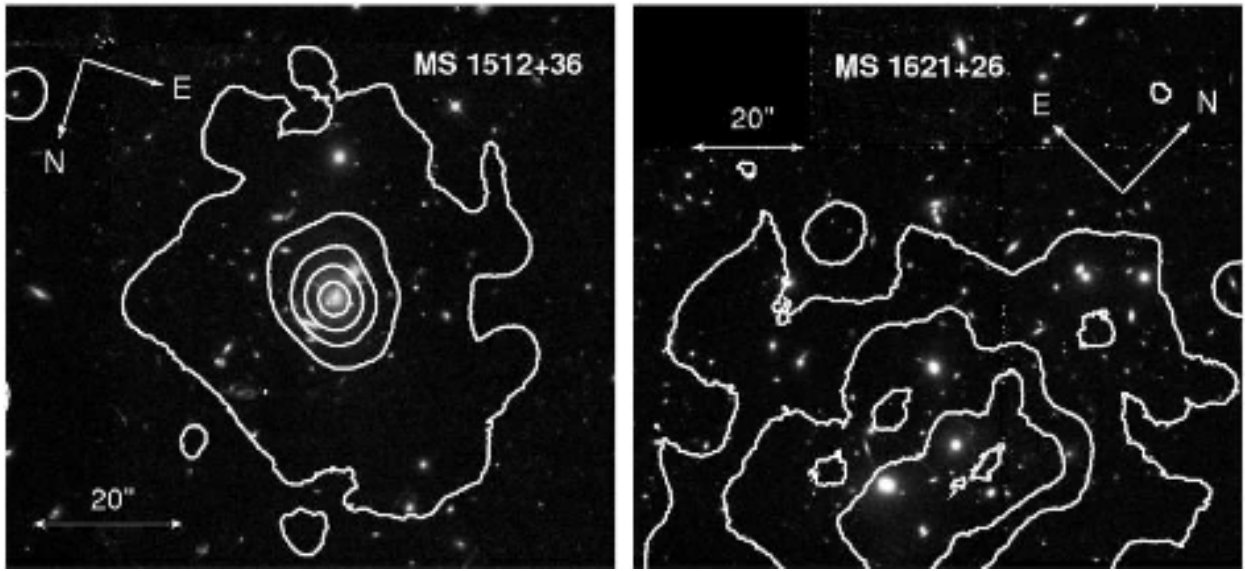}
\caption{{\bf{Optical Images.}}  X-ray contours are overlaid on optical images of clusters in our sample.  HST images of Abell 2390, MS0016+16, MS0302+16, MS0440.5+02, MS0451-03, MS1358+62, MS1455+22, MS1512+36, and MS1621+26 were retrieved from the MAST archive, with respective filters F814W, F555W, F814W, F702W, F702W, F814W, F606W, F555W, and F814W.  Gunn $r$ band (to $r\sim24$) CFHT images for the remaining five clusters (MS0451+02, MS0839+29, MS0906+11, MS1006+12, and MS1008-12) are from the CNOC1 survey~\citep{cnoccat}.  X-ray countours were created from adaptively smoothed Chandra 0.6-7.0 keV flux images, and all have linear values except for three of the cooling core clusters (Abell 2390, MS0440+02, and MS1455+22) which are overlayed with logarithmic X-ray contours.  Note the significant amount of substructure present in MS0451+02.\label{fig1}}
\end{figure}

\clearpage
\begin{figure}
\centerline{\includegraphics[width=2.5in, angle=90]{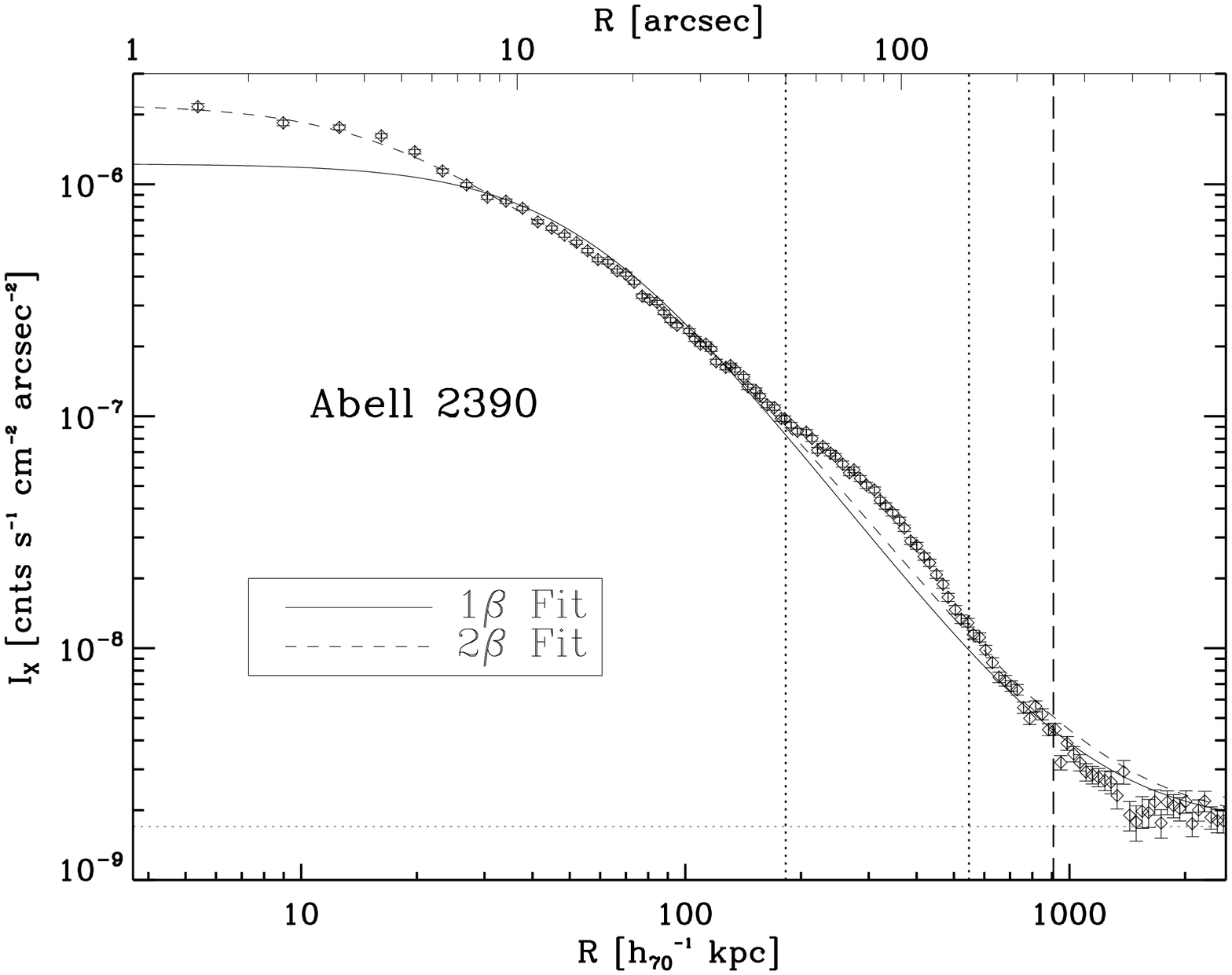}
\includegraphics[width=2.5in, angle=90]{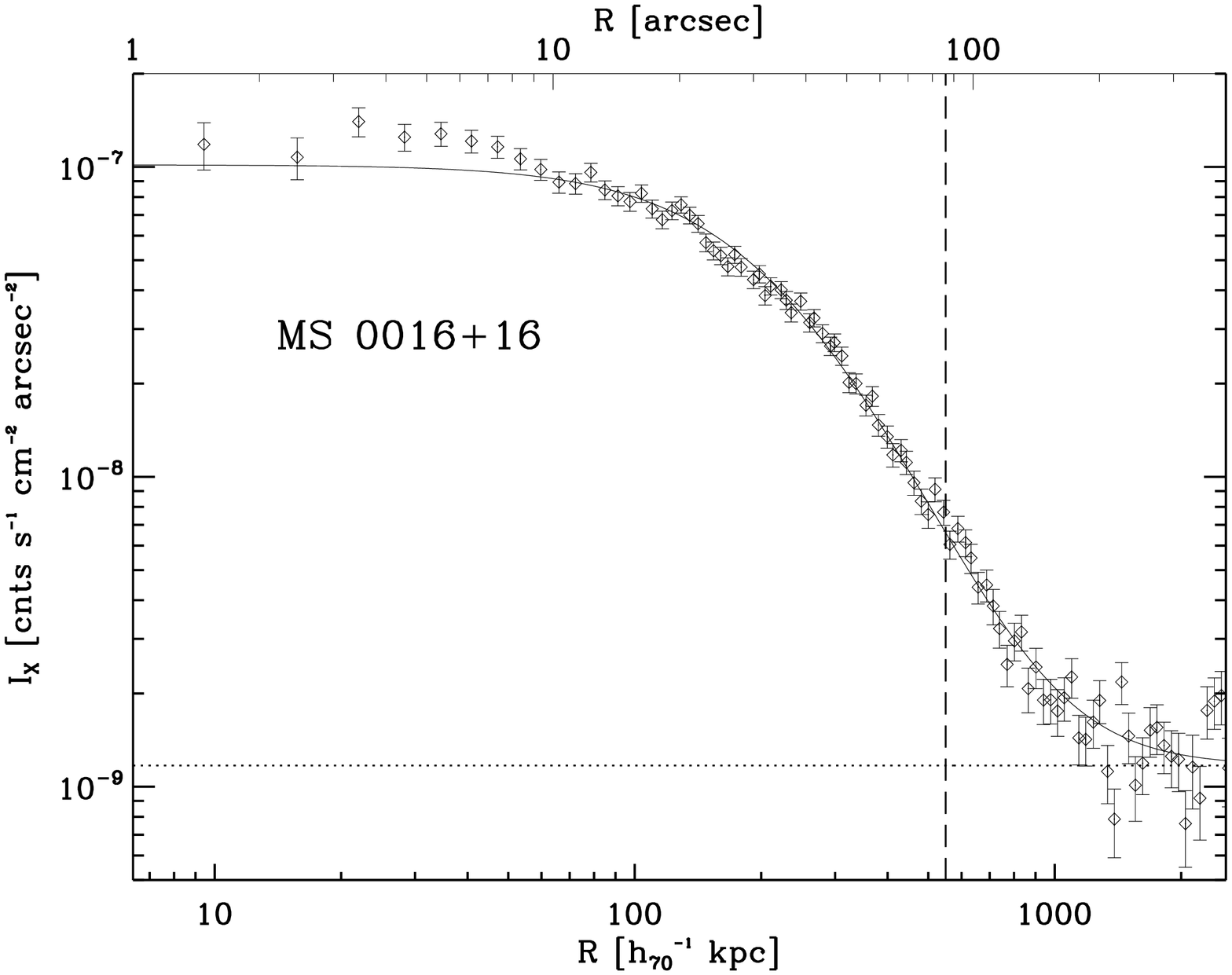}}
\centerline{\includegraphics[width=2.5in, angle=90]{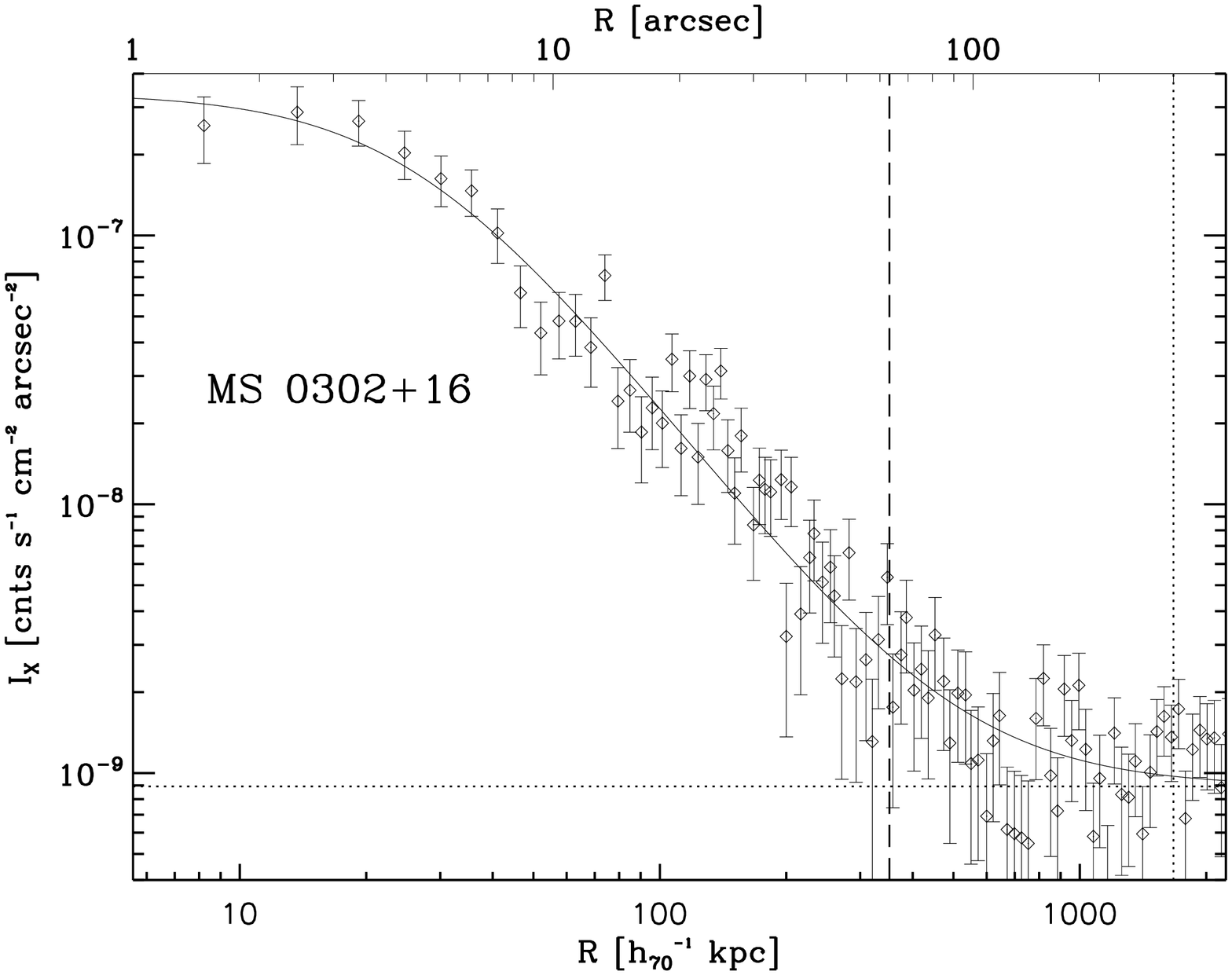}
\includegraphics[width=2.5in, angle=90]{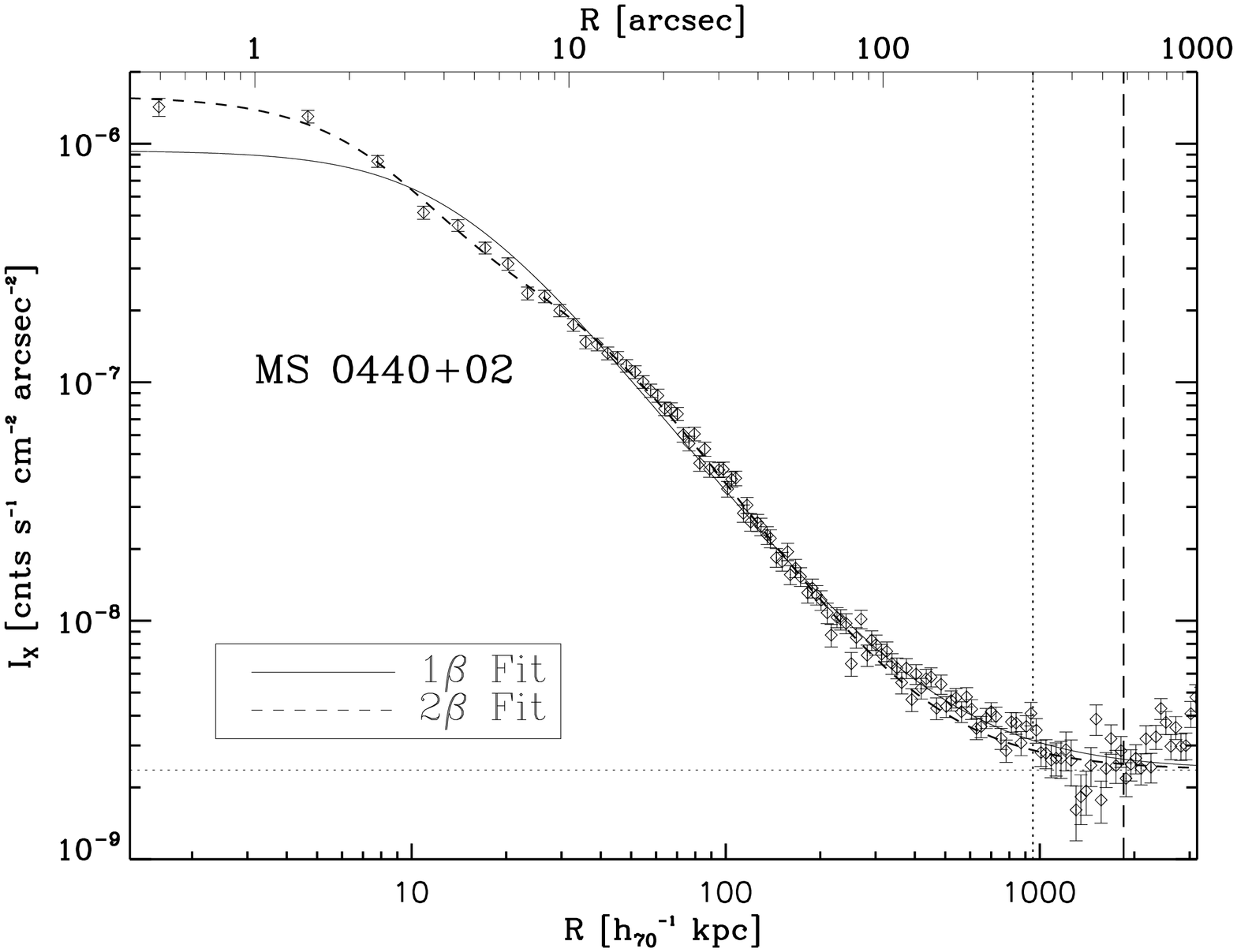}}
\centerline{\includegraphics[width=2.5in, angle=90]{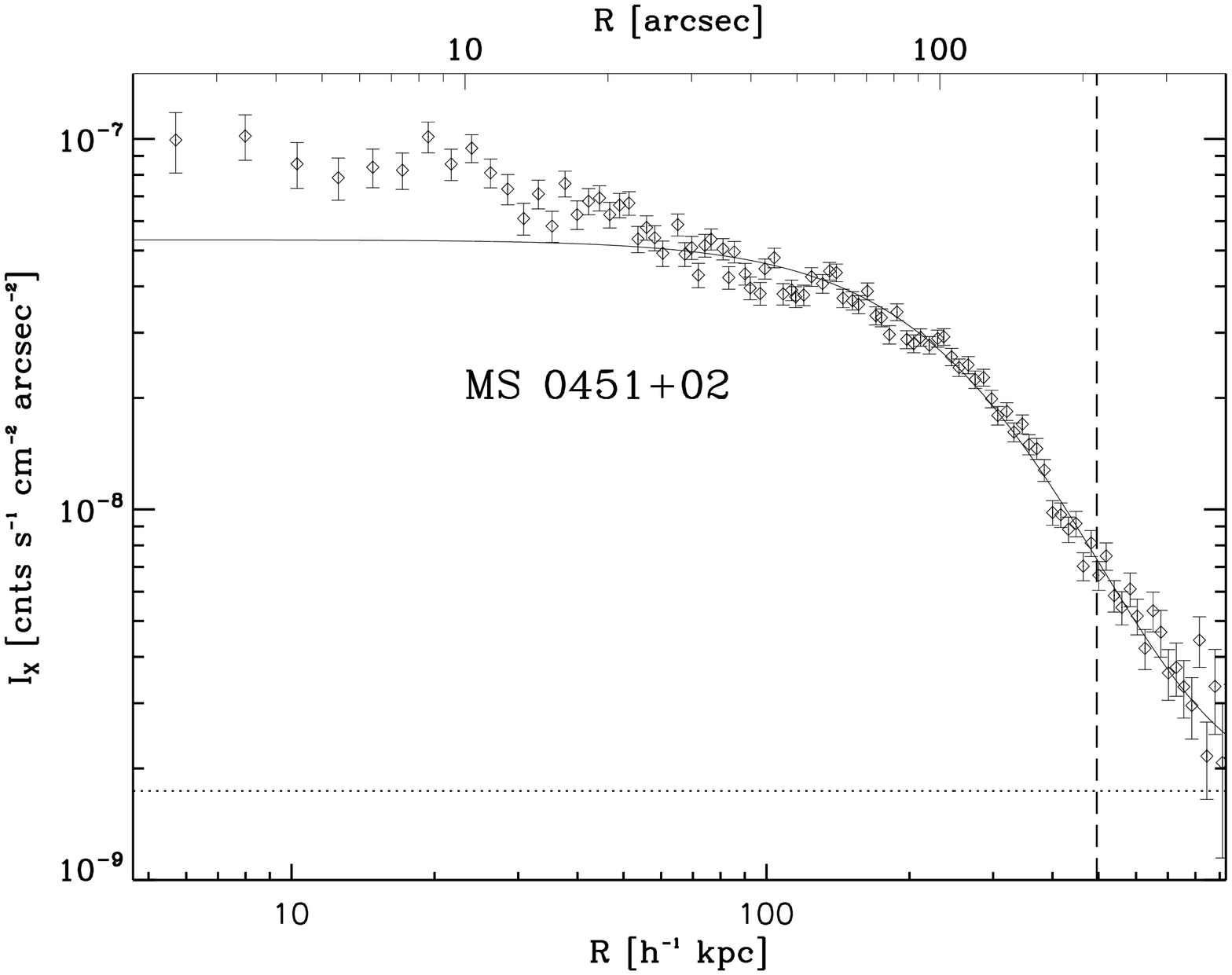}
\includegraphics[width=2.5in, angle=90]{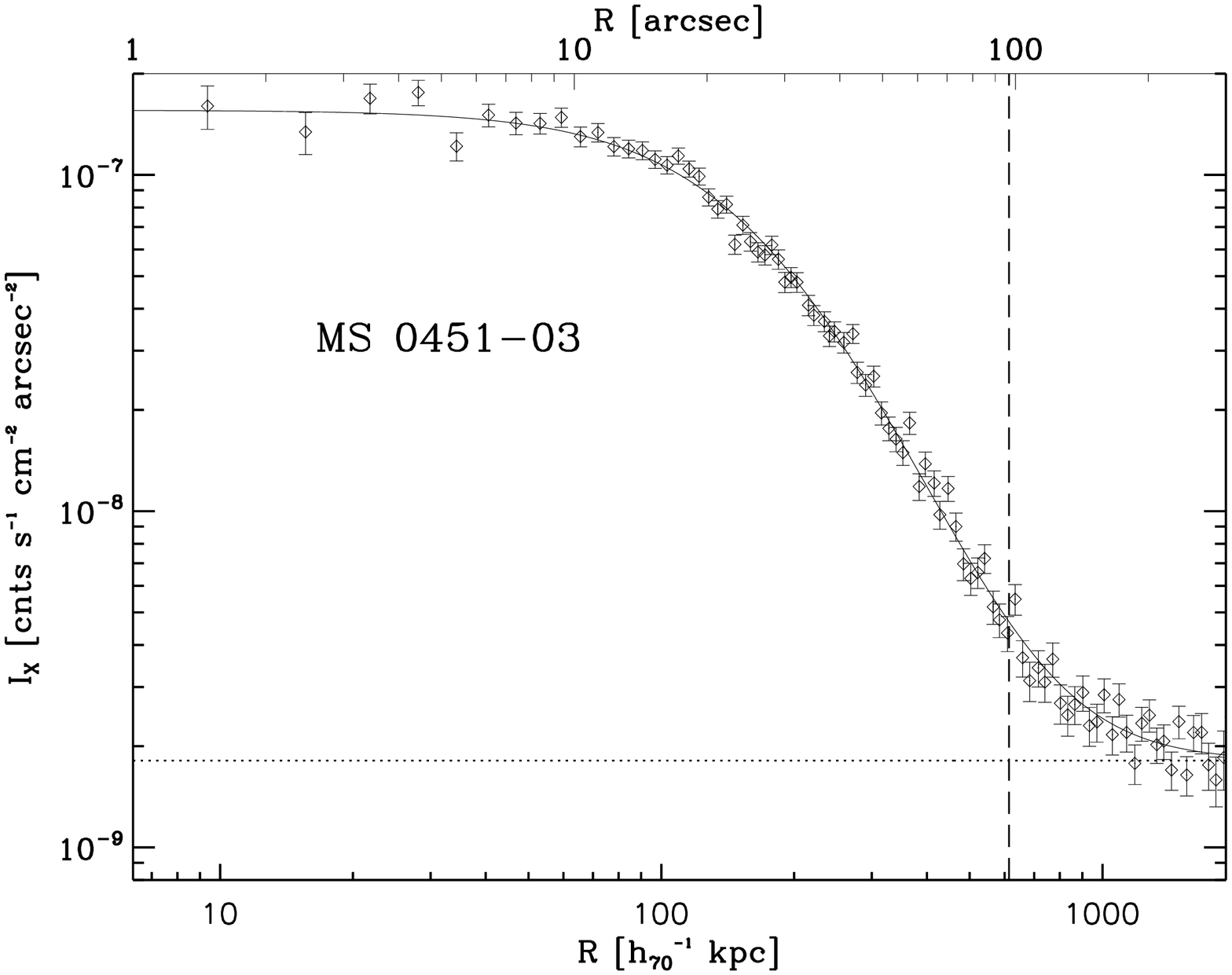}}
\end{figure}
\begin{figure}
\centerline{\includegraphics[width=2.5in, angle=90]{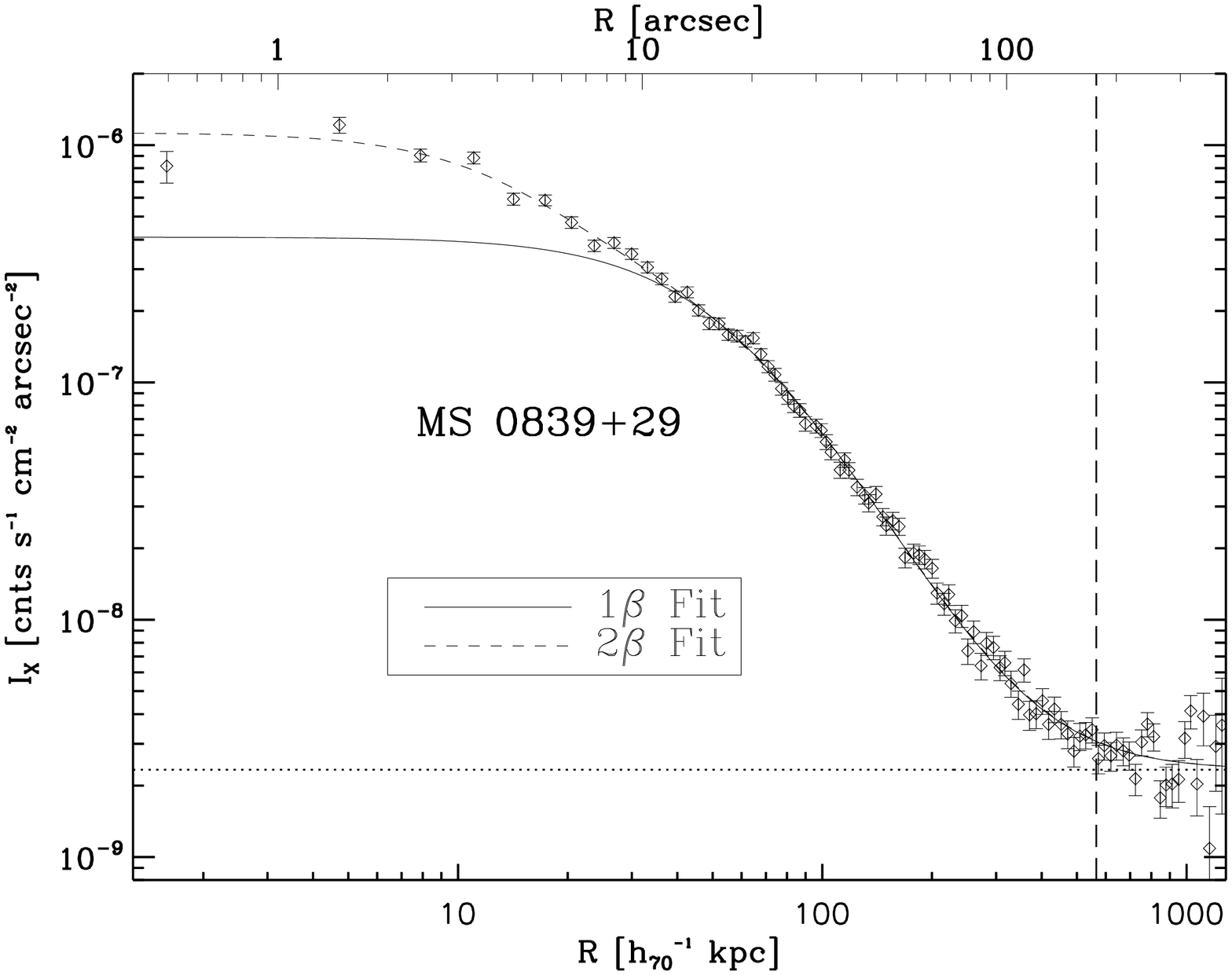}
\includegraphics[width=2.5in, angle=90]{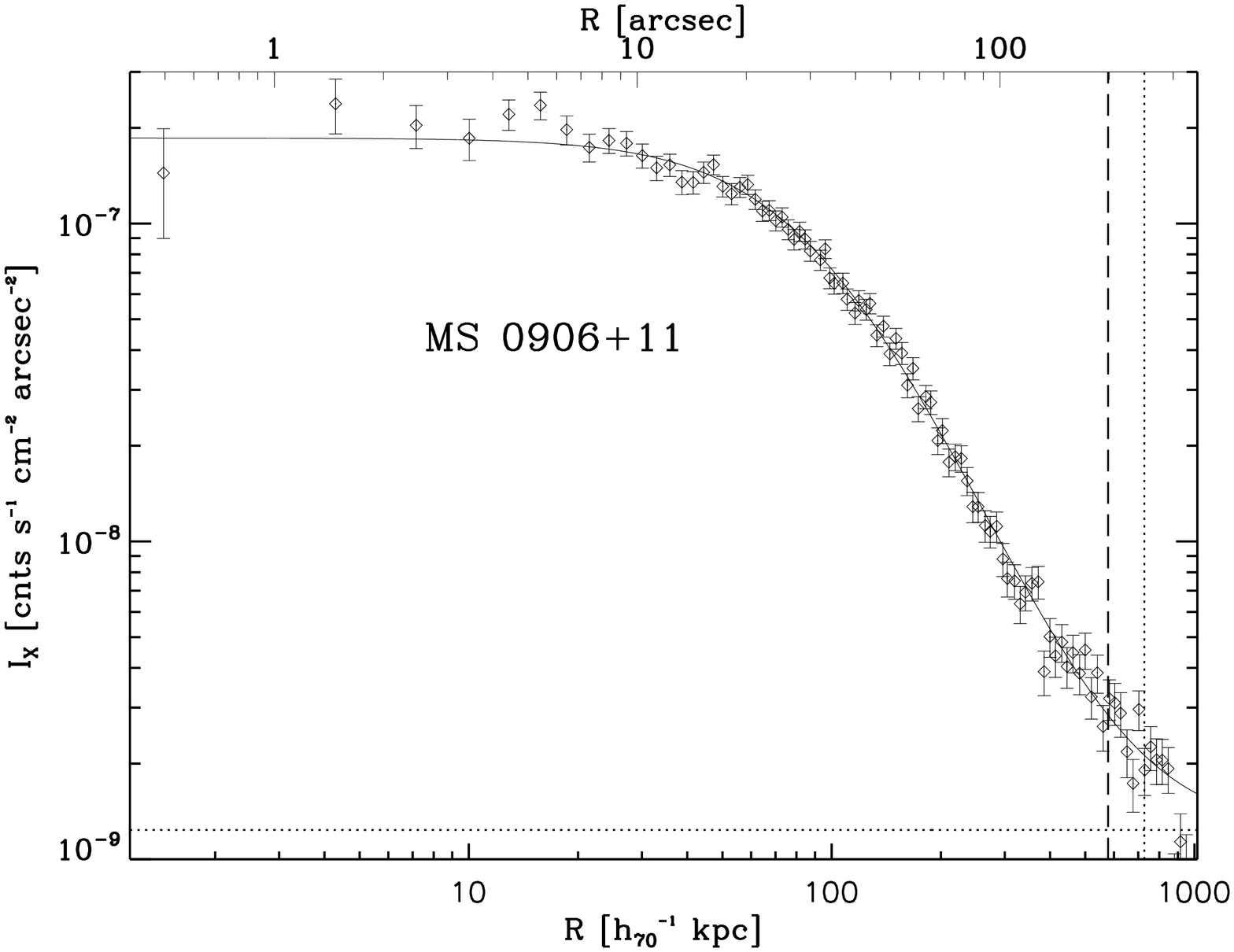}}
\centerline{\includegraphics[width=2.5in, angle=90]{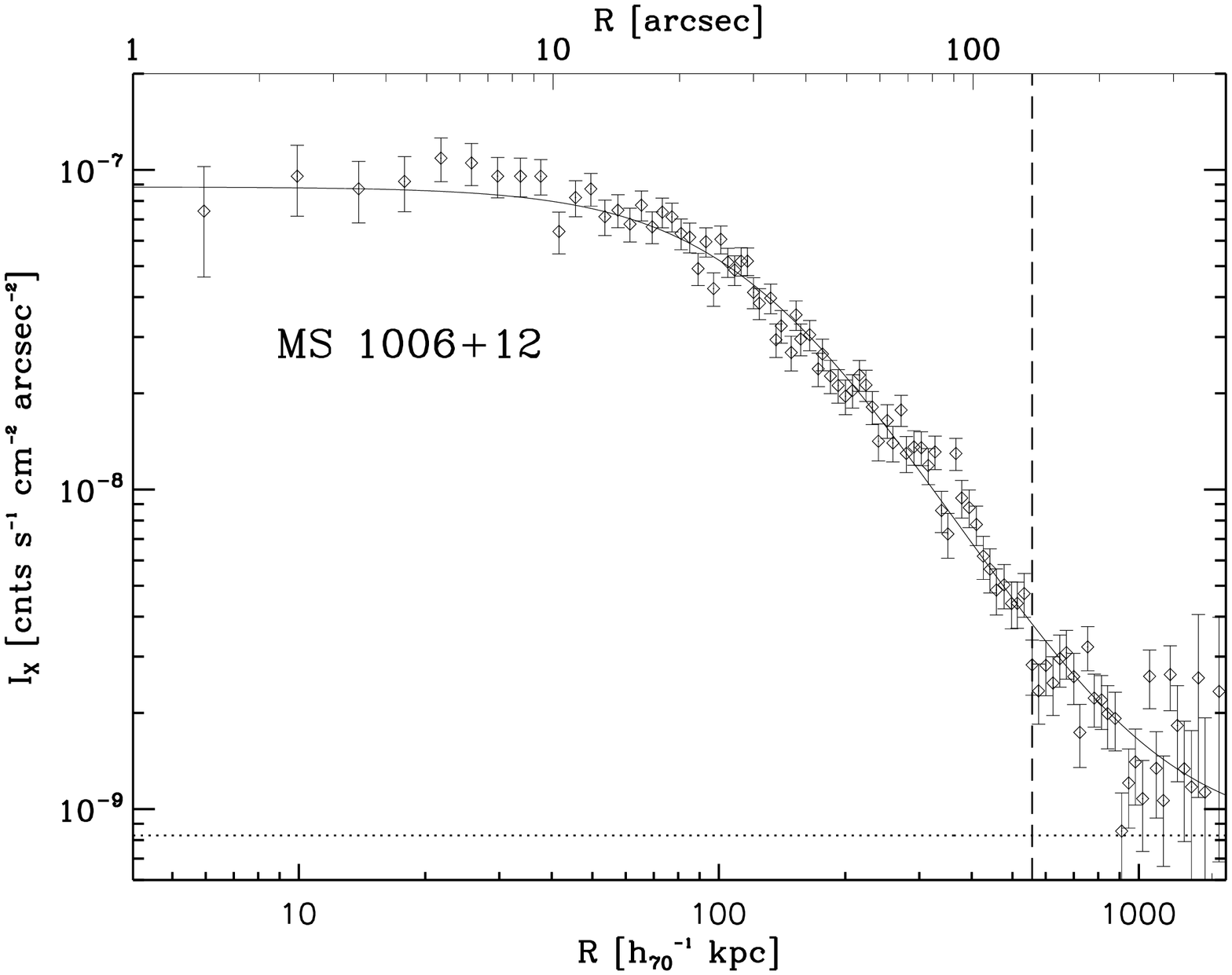}
\includegraphics[width=2.5in, angle=90]{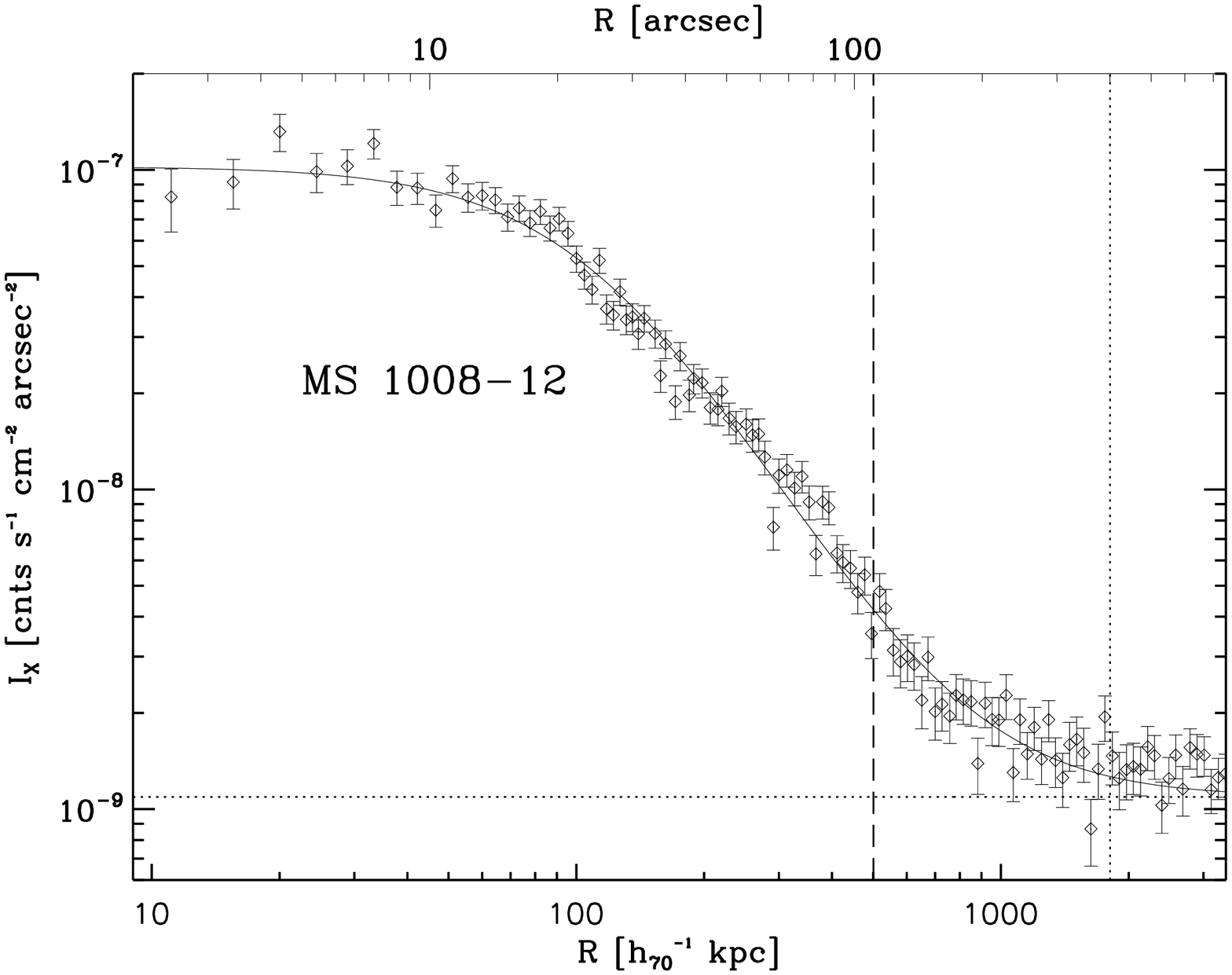}}
\centerline{\includegraphics[width=2.5in, angle=90]{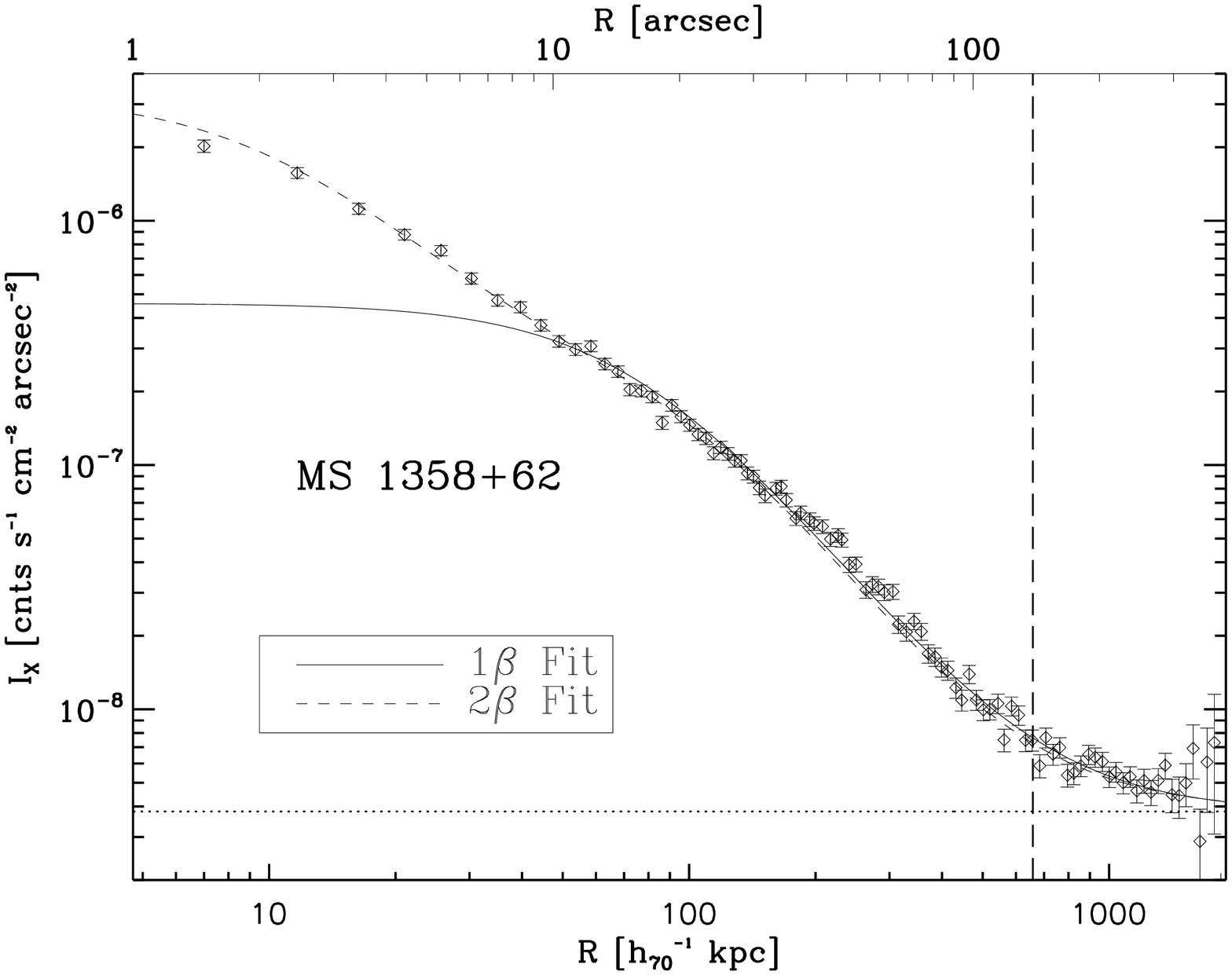}
\includegraphics[width=2.5in, angle=90]{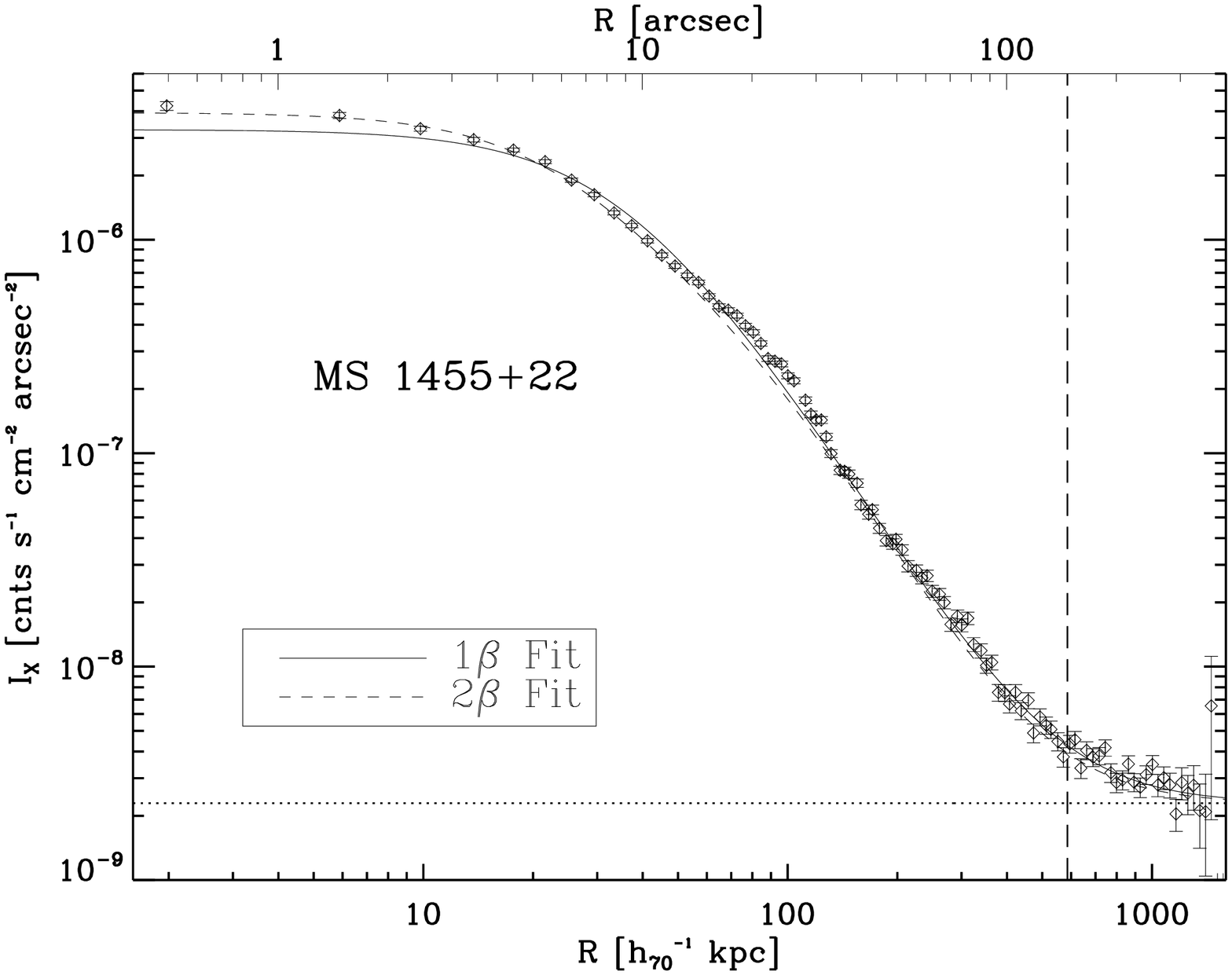}}
\end{figure}
\begin{figure}
\centerline{\includegraphics[width=2.5in, angle=90]{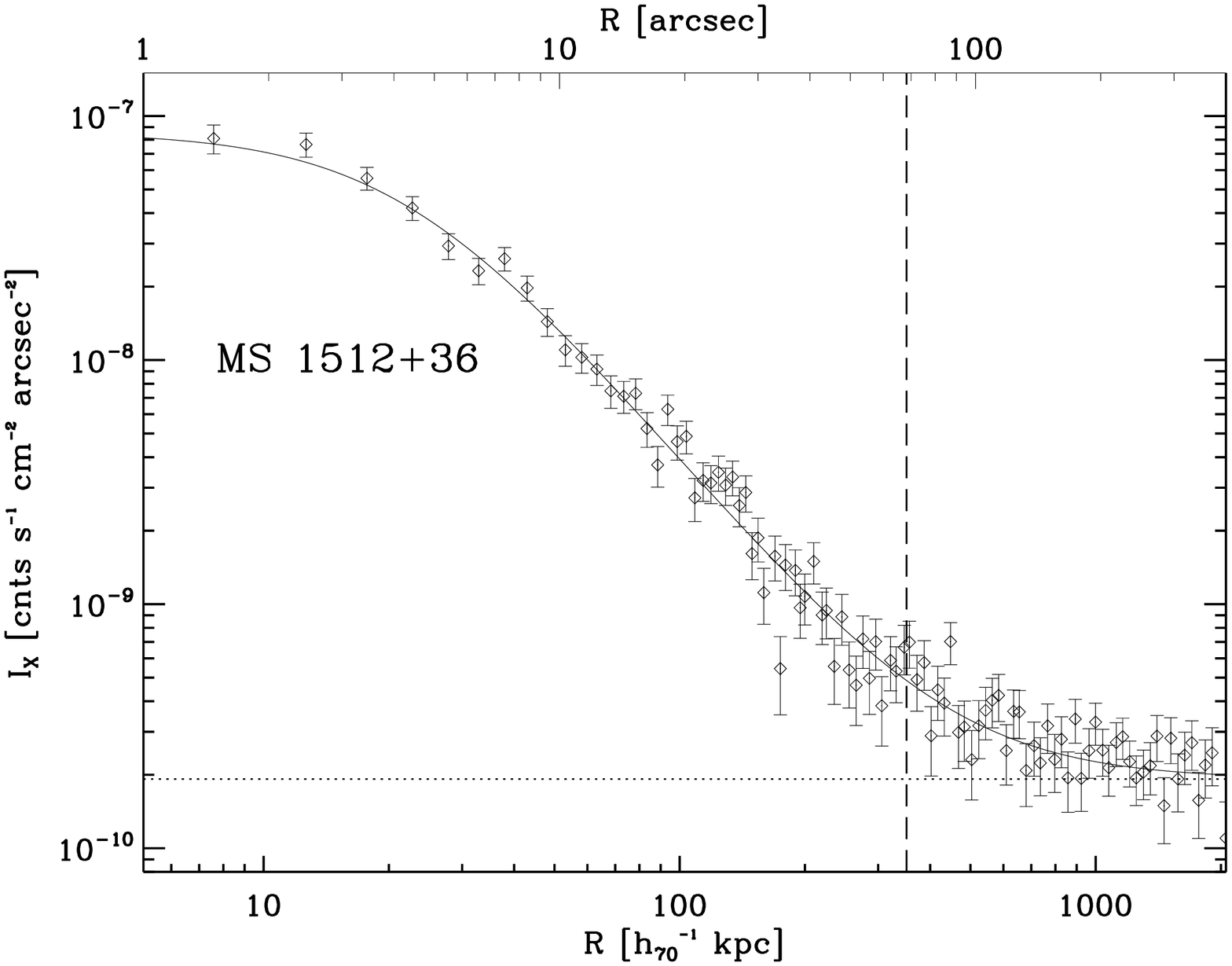}
\includegraphics[width=2.5in, angle=90]{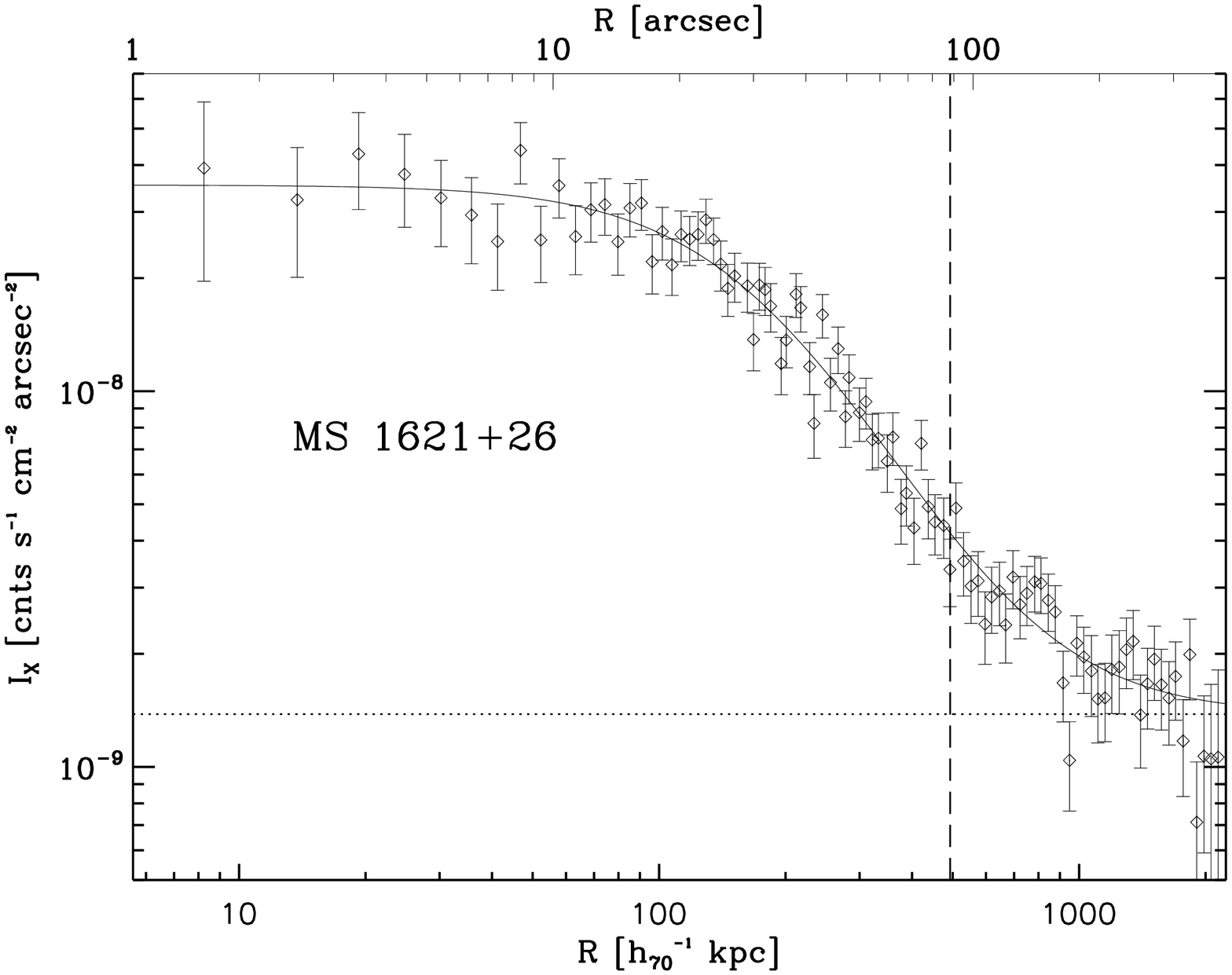}}
\caption{{\bf{Surface Brightness Profiles.}} \label{fig2} Radial surface brightnesses of CNOC clusters calculated in $1\arcsec$ annular bins using 0.6-7.0 keV data.  Horizontal dotted lines denote the best-fitting background of the model.  Pairs of vertical dotted lines indicate regions of the profile which were excluded from fitting, while a single vertical dotted line denotes the radius to which fitting was performed.  Solid lines describe the best fitting single $\beta$ model for each cluster, while short dashed lines indicate double $\beta$ model fits to the data.  Long dashed vertical lines represent $\rm{R}_{2500}$.  Cluster emission is well described by either a single or double $\beta$ model except in the case of MS0451+02, which exhibits a significant amount of substructure.  Seemingly high values of reduced $\chi^2$ are a consequence of fitting high resolution data with small error bars, such that even small deviations from symmetry will drive $\chi^2$ up significantly.}
\end{figure}

\clearpage

\begin{figure}
\includegraphics{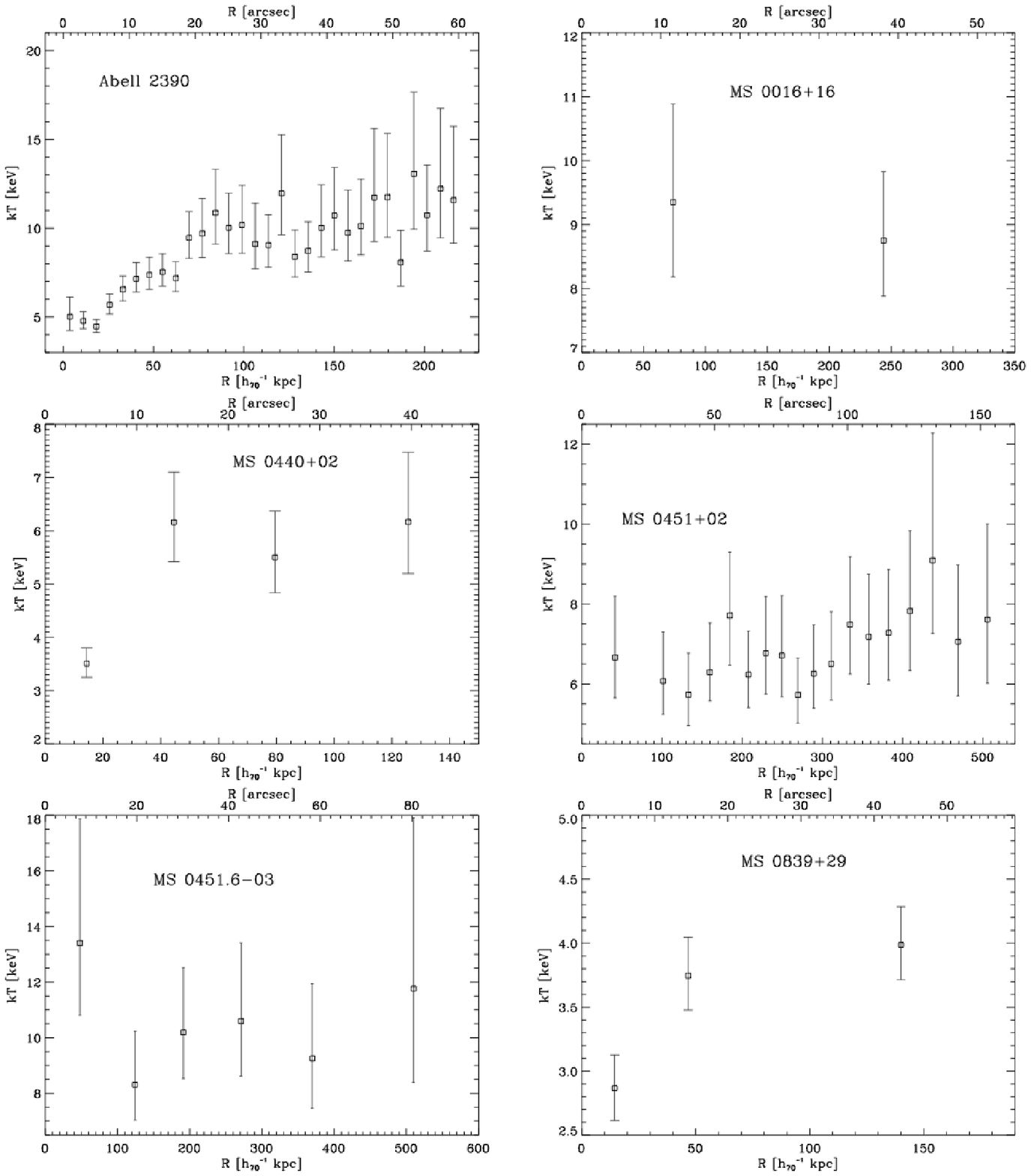}

\end{figure}
\begin{figure}
\includegraphics{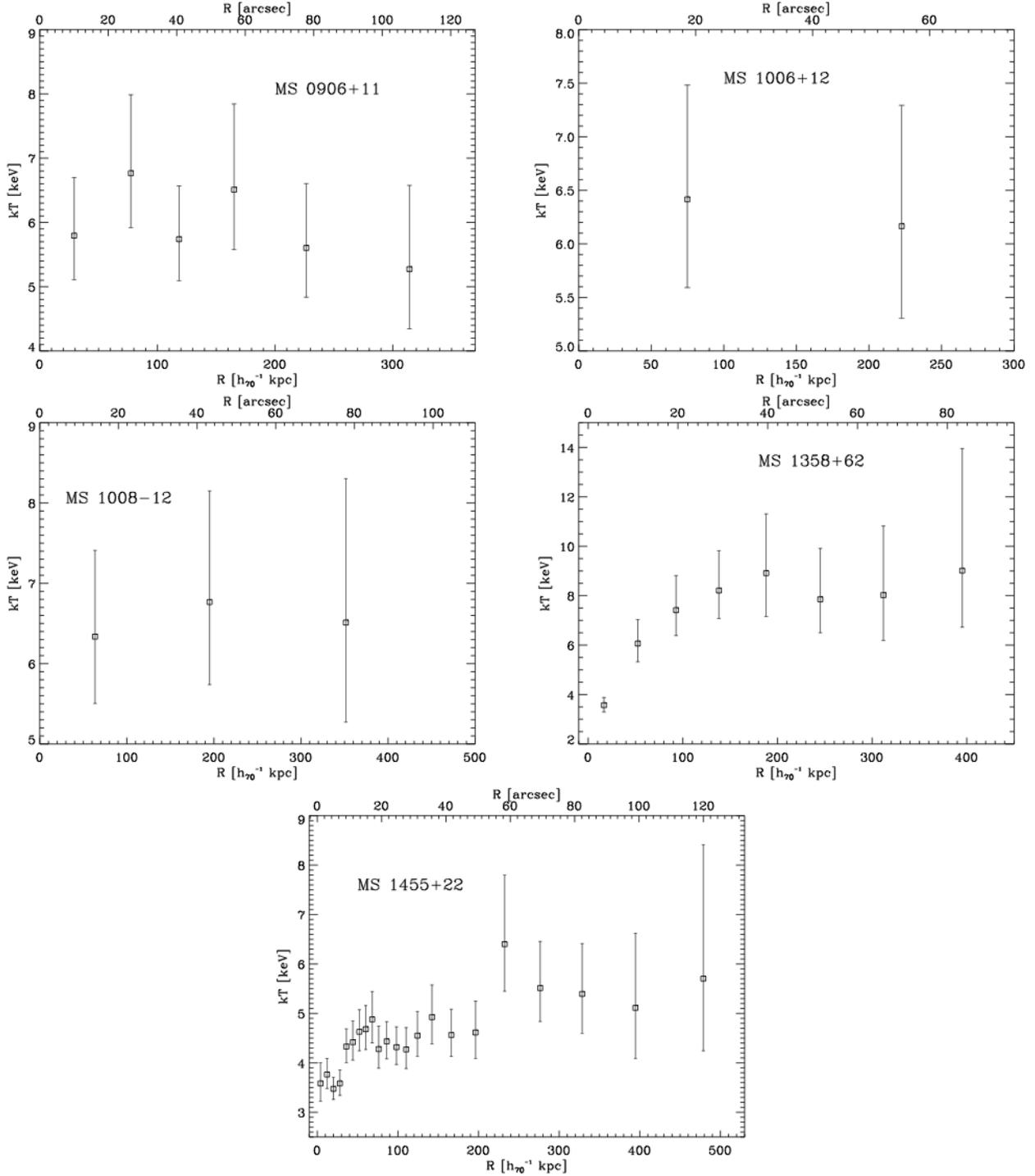}

\caption{{\bf{Temperature Profiles.}} \label{fig3} Radial temperature profiles determined by fitting, in XSPEC, single temperature spectral models to spectra extracted from circular annuli selected to include a minimum of 2500 counts in the 0.6-7.0 keV band.  Error bars represent 90\% confidence limits.  Indications of cool cores are seen in Abell 2390, MS0440+02, MS0839+29, MS1358+62, and MS1455+22.}
\end{figure}

\clearpage

\begin{figure}
\centerline{\includegraphics[angle=90,width=4.5in]{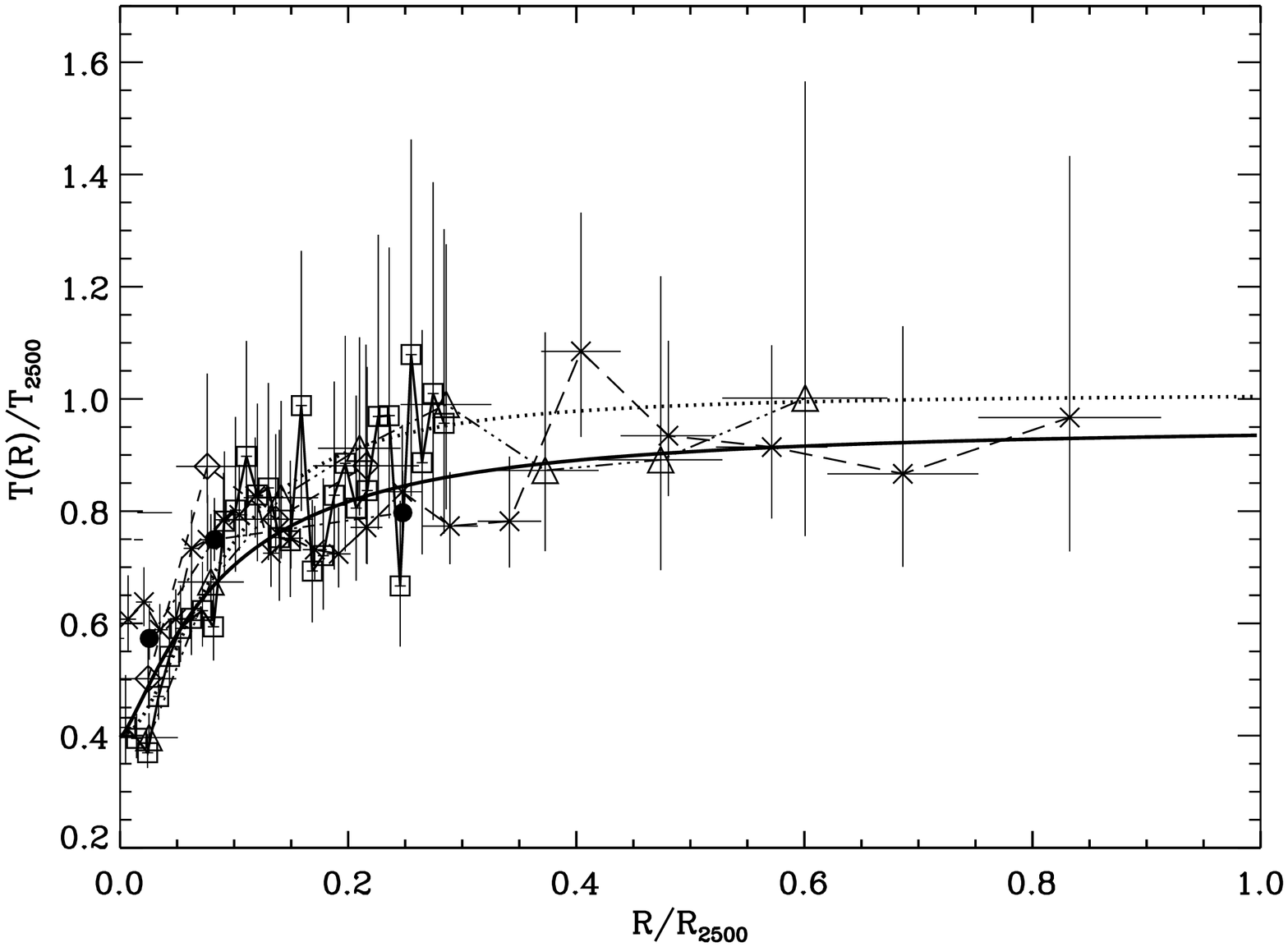}}
\centerline{\includegraphics[angle=90,width=4.5in]{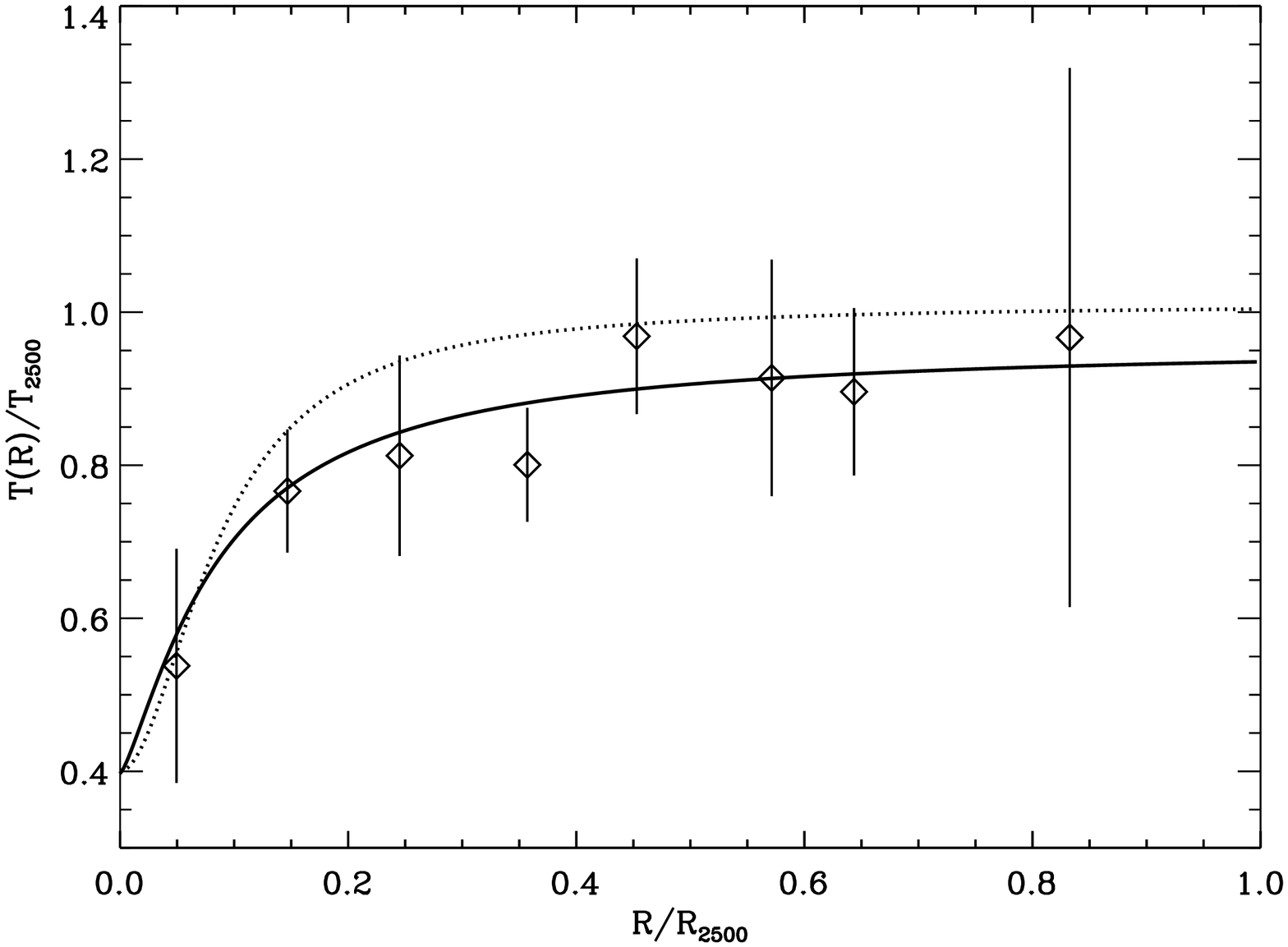}}
\caption{{\bf{Universal Temperature Profiles.}} \label{fig4} Top - Individual temperature profiles of the five cooling core clusters in this sample. Abell 2390: squares and solid line, MS0440+02: diamonds and dashed line, MS 0839+29: filled circles and dot-dash line, MS1358+62: triangles and dot-dot-dot-dash line, MS455+22: x-symbols and long dashed line.  The best fit to the data is overlayed as a thick solid line, while the dotted line indicates the best fit found by~\citet{allen01}.  Bottom - Weighted average temperature profile of the combined five clusters in 0.1 $\rm{R}/\rm{R}_{2500}$ bins, with best fitting models overlayed as described above.}
\end{figure}

\clearpage

\begin{figure}
\centerline{\includegraphics[angle=90,width=6.5in]{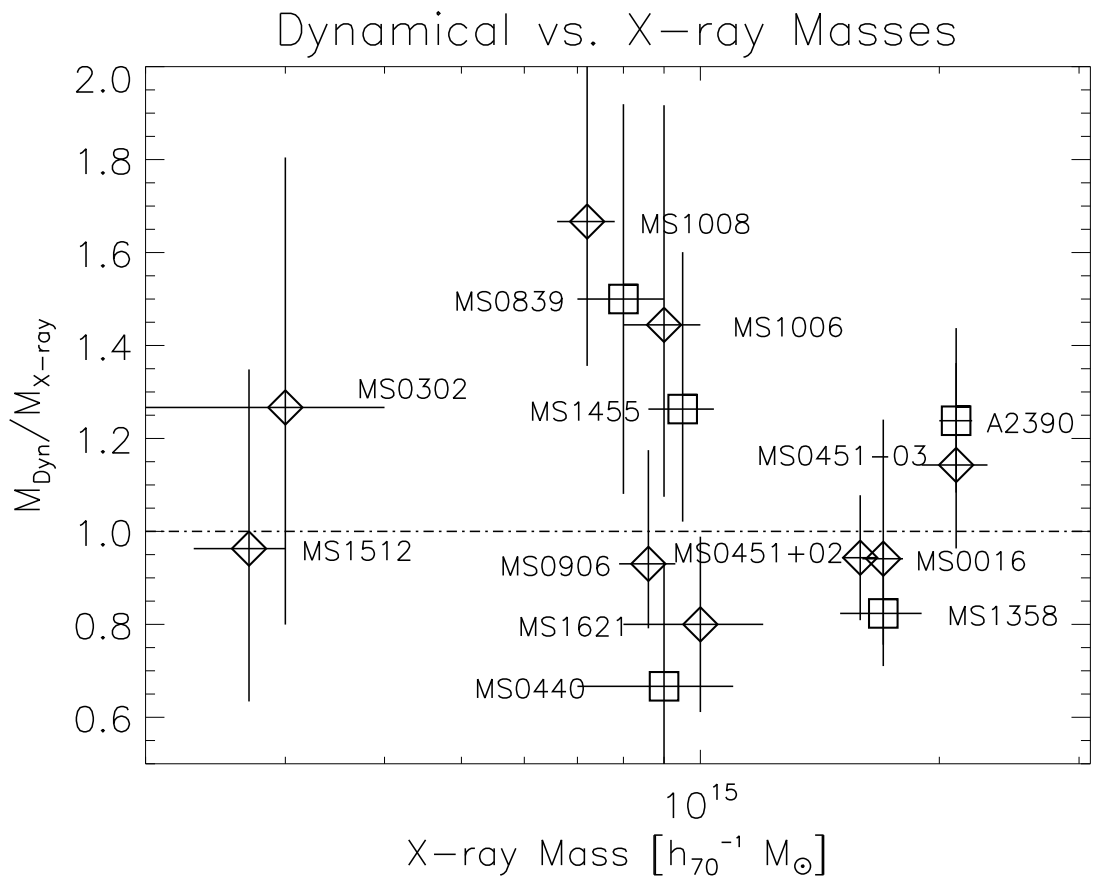}}
\caption{{\bf{Dynamical vs. X-ray Masses.}} \label{fig5} X-ray mass estimates are plotted against the ratio $\rm{M}_{\rm{Dyn}}/\rm{M}_{\rm{X-ray}}$.  The dot-dash line indicates a mass ratio of 1.0.  Squares represent cooling core clusters.  Error bars denote 68\% confidence limits.}
\end{figure}

\clearpage

\begin{figure}
\centerline{\includegraphics[angle=90,width=6.5in]{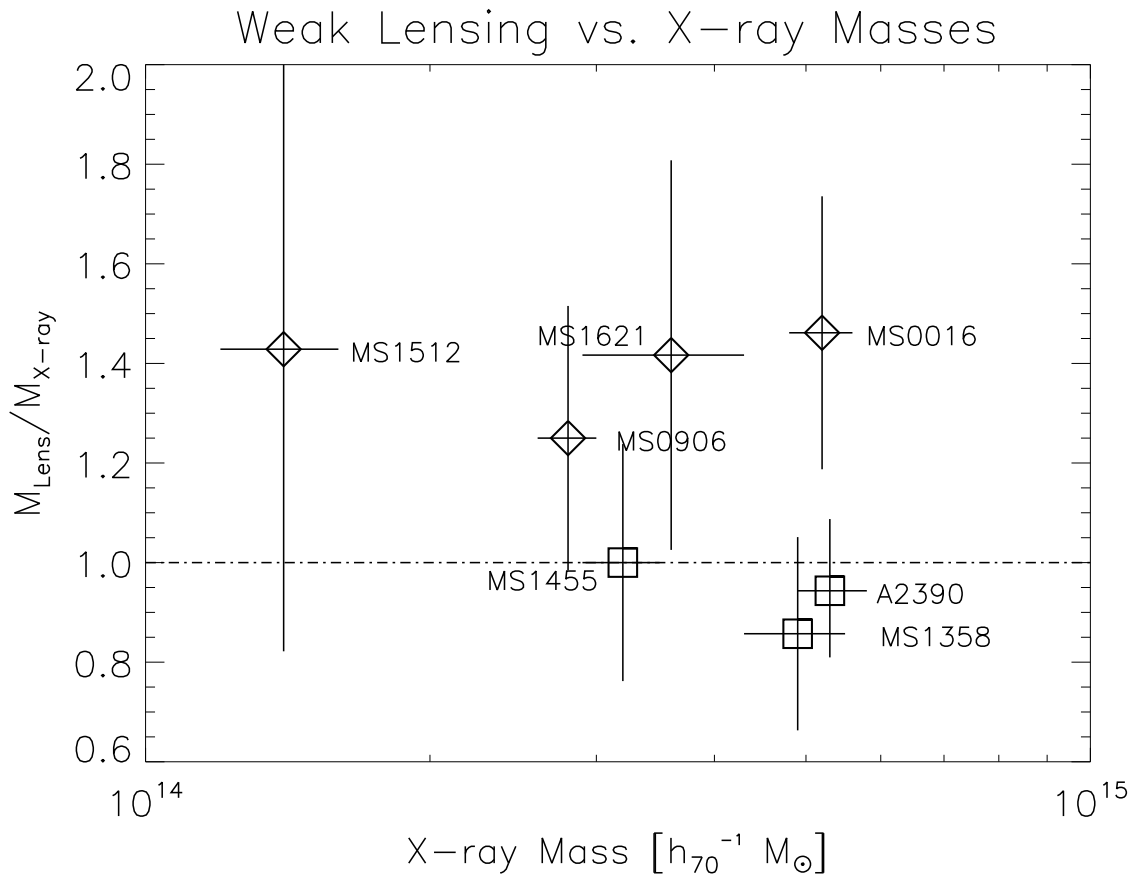}}
\caption{{\bf{Weak Lensing vs. X-ray Masses.}} \label{fig6}  X-ray mass estimates are plotted against the ratio $\rm{M}_{\rm{Lens}}/\rm{M}_{\rm{X-ray}}$.  The dot-dash line represents a mass ratio of 1.0.  Squares indicate cooling core clusters, and error bars denote 68\% confidence limits.  Though the distribution appears assymmetric, it is not statistically significant.}
\end{figure}

\clearpage

\begin{figure}
\centerline{\includegraphics[angle=90,width=6.5in]{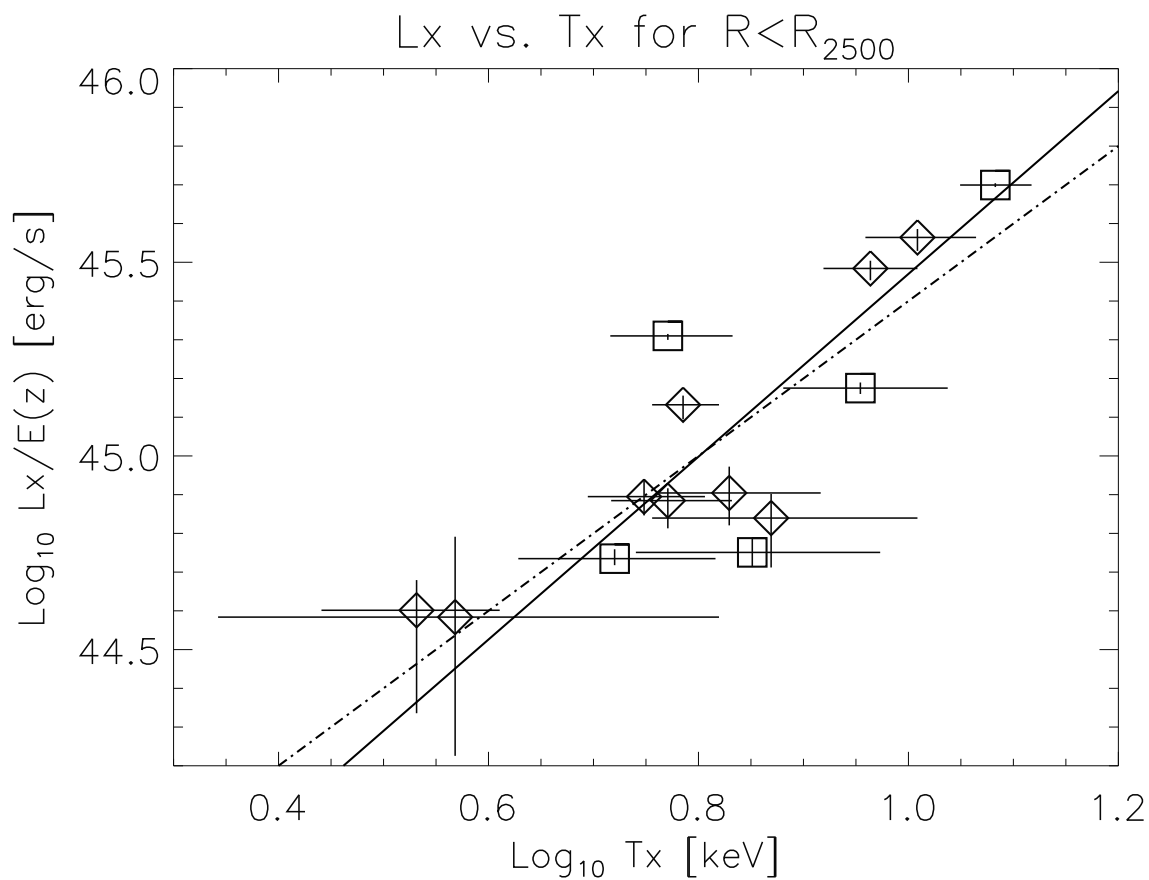}}
\caption{{\bf{The $L_X-T_X$ Relationship.}} \label{fig7}   X-ray temperatures are plotted against cosmologically corrected unabsorbed bolometric luminosities within $\rm{R}_{2500}$.  Squares represent clusters with cool cores.  The solid line indicates the best fitting relationship, with a slope of $2.4\pm{0.2}$.  The dash-dot line illustrates a slope of 2.0.  Error bars indicate 90\% confidence limits.}
\end{figure}

\clearpage

\begin{figure}
\centerline{\includegraphics[angle=90,width=6.5in]{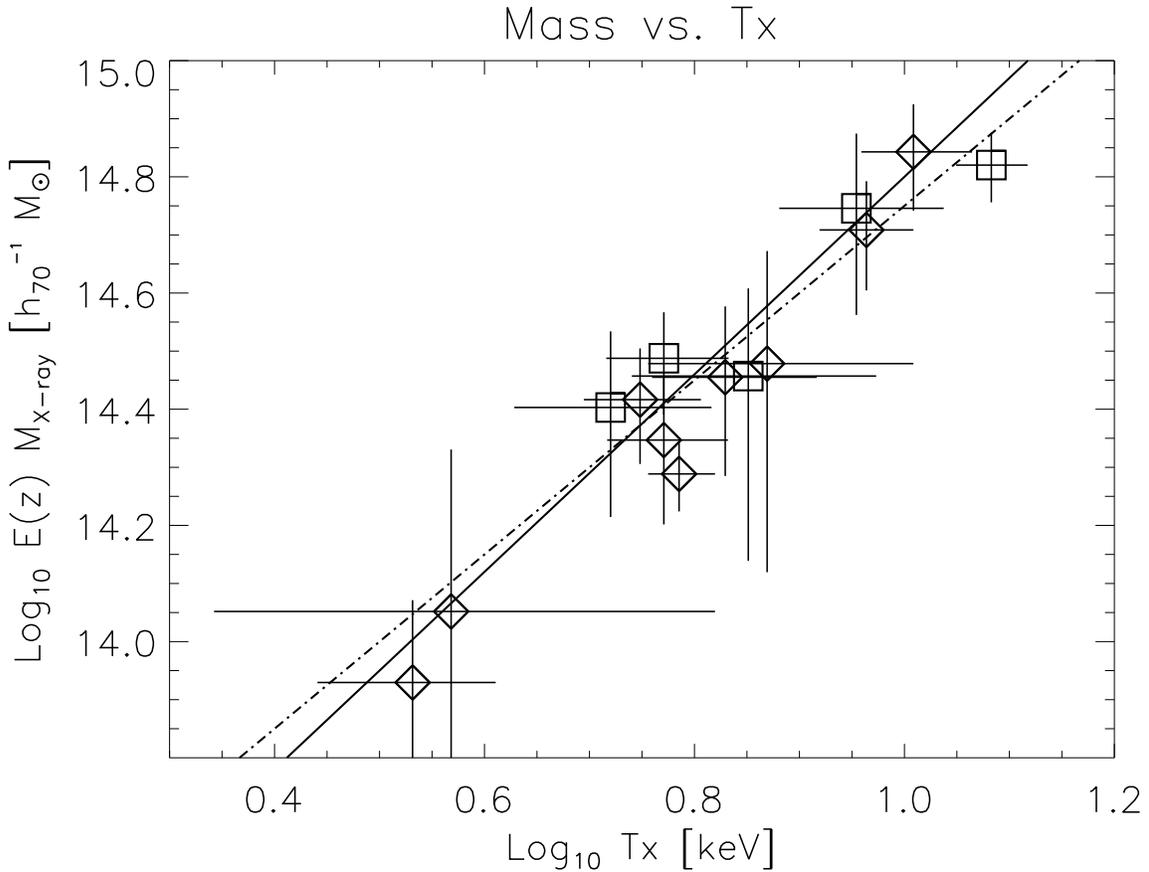}}
\caption{{\bf{The Mass-$T_X$ Relationship.}} \label{fig8}   X-ray temperatures are plotted against X-ray derived masses within $\rm{R}_{2500}$.  Squares represent clusters with cool cores.  The solid line indicates the best fitting relationship, with a slope of $1.7\pm{0.1}$, while the dash-dot line represents a slope of 1.5.  Error bars indicate 90\% confidence limits.}
\end{figure}

\clearpage

\begin{figure}
\centerline{\includegraphics[angle=90,width=6.5in]{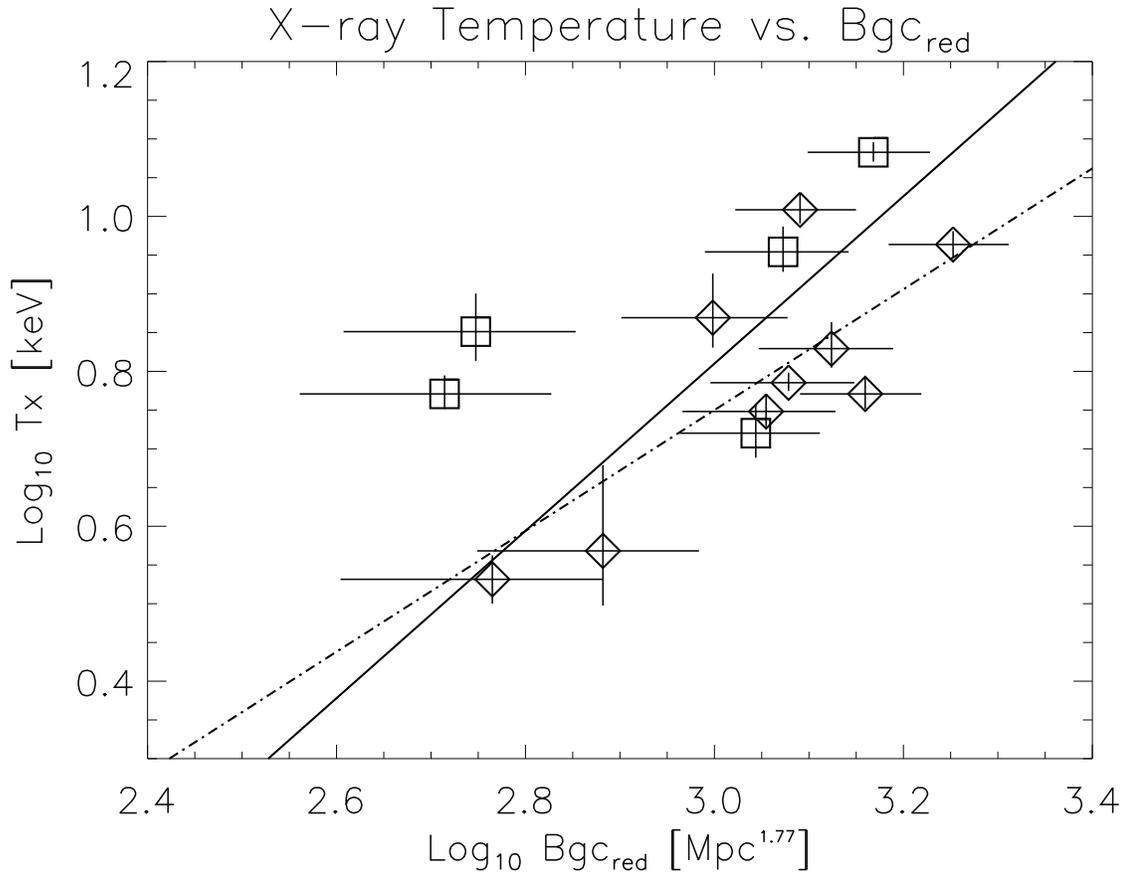}}
\caption{{\bf{$T_X$ vs. $\rm{B}_{gc,red}$.}} \label{fig9}   A log-log plot of $T_X$ vs. $\rm{B}_{gc,red}$ for the clusters in our sample.  Error bars represent 68\% confidence intervals.  The solid line indicates the best fitting relationship between these two parameters (Section~\ref{s:or}), while the dash-dot line illustrates the previous results of~\citet{yee03}.  Squares denote clusters which exhibit significant cool cores.}
\end{figure}

\clearpage
\begin{figure}
\centerline{\includegraphics[angle=90,width=6.5in]{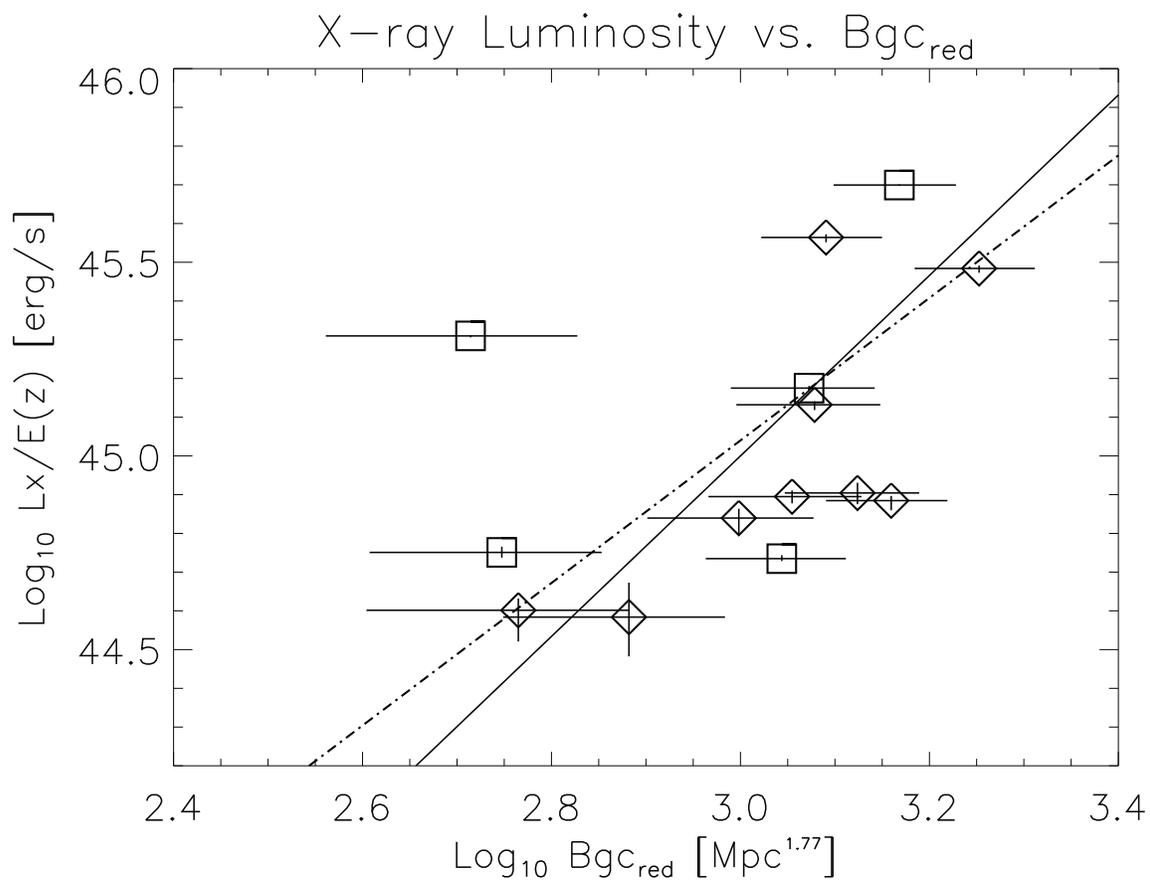}}
\caption{{\bf{$L_X$ vs. $\rm{B}_{gc,red}$.}} \label{fig10} A log-log plot of $L_X$ vs. $\rm{B}_{gc,red}$ for the clusters in our sample.  Error bars represent 68\% confidence intervals.  The solid line indicates the best fitting relationship (Section~\ref{s:or}), and the dash-dot line denotes the~\citet{yee03} result.  Squares again denote clusters which exhibit significant cool cores.}
\end{figure}

\clearpage
\begin{figure}
\centerline{\includegraphics[angle=90,width=6.5in]{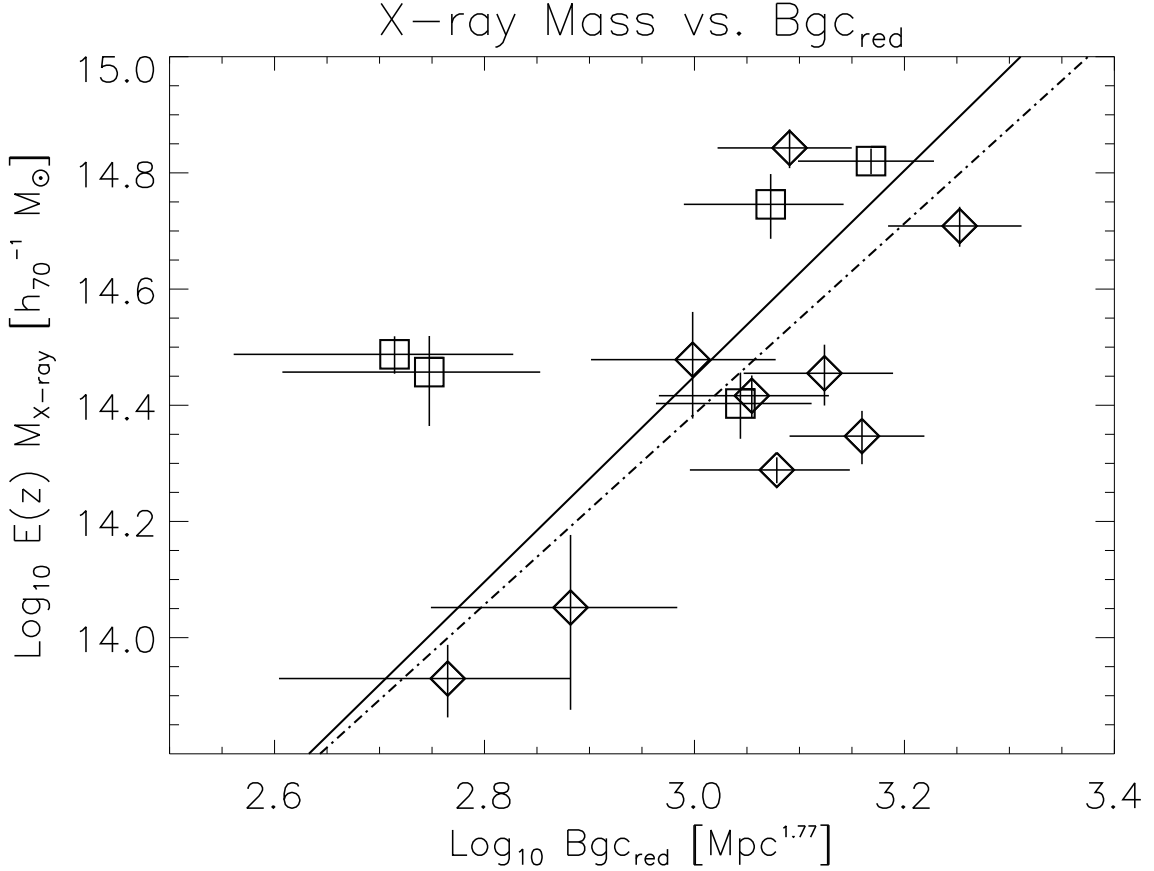}}
\caption{{\bf{X-ray Mass vs. $\rm{B}_{gc,red}$.}} \label{fig11}   A log-log plot of X-ray mass estimates vs. $\rm{B}_{gc,red}$ for the clusters in our sample.  Error bars represent 68\% confidence intervals.  The solid line represents the best fitting relationship between these two parameters (Section~\ref{s:or}).  The dash-dot line was produced from the~\citet{yee03} results, scaling their intercept by the average of $\rm{M}_{2500}/\rm{M}_{200}$ for this sample.  Squares denote cooling core clusters.}
\end{figure}

\clearpage


\begin{deluxetable}{ccccccccc}
\tabletypesize{\small}
\tablecolumns{9}
\tablewidth{0pt}
\tablecaption{Cluster Sample\label{table1}}
\tablehead{
\multicolumn{2}{c}{Cluster} & 
\colhead{{\rm{z}}} &
\multicolumn{2}{c}{Centroid}& 
\colhead{1$\arcsec$} & 
\colhead{obsid} &
\colhead{Array} & 
\colhead{Chandra Exposure\tablenotemark{a}} \\
\multicolumn{2}{c}{} & 
\colhead{} &
\colhead{[RA]}&
\colhead{[Dec]}&
\colhead{[$h_{70}^{-1}$ kpc]} &
\colhead{} & 
\colhead{} & 
\colhead{[seconds]} 
}
\startdata
\multicolumn{2}{l}{Abell 2390} & 0.2279 & 21:53:36.794& +17:41:41.85& 3.65  & 4193 & ACIS-S & 89624\\
\multicolumn{2}{l}{MS 0015.9+1609} &  0.5466& 00:18:33.64& +16:26:11.2& 6.39   & 520 & ACIS-I & 69235 \\
\multicolumn{2}{l}{MS 0302.7+1658} &  0.4246& 03:05:31.72& +17:10:01.5& 5.57   & 525 &ACIS-I & 11764 \\
\multicolumn{2}{l}{MS 0440.5+0204} & 0.1965& 04:43:09.974 & +02:10:18.01& 3.17   & 4196 &ACIS-S & 45104 \\
\multicolumn{2}{l}{MS 0451.5+0250} &  0.2010&  \nodata& \nodata& 3.31  &4215 & ACIS-I & 66275 \\
\multicolumn{2}{l}{MS 0451.6-0305} &  0.5392&  04:54:11.19& -03:00:52.2& 6.34  & 902 &ACIS-S & 43652  \\
\multicolumn{2}{l}{MS 0839.8+2938} &  0.1928& 08:42:55.999& +29:27:25.45& 3.21   & 2224 &ACIS-S & 31383 \\
\multicolumn{2}{l}{MS 0906.5+1110} &  0.1709&  09:09:12.81& +10:58:31.7& 2.91  & 924 &ACIS-I & 31392 \\
\multicolumn{2}{l}{MS 1006.0+1202} &  0.2605& 10:08:47.56& +11:47:34.0&4.03   & 925 &ACIS-I & 25590 \\
\multicolumn{2}{l}{MS 1008.1-1224} &  0.3062&10:10:32.44 & -12:39:41.4& 4.52   & 926 &ACIS-I & 37376 \\
\multicolumn{2}{l}{MS 1358.4+6245} &  0.3290& 13:59:50.640& +62:31:04.20& 4.74   & 516 &ACIS-S & 53055 \\
\multicolumn{2}{l}{MS 1455.0+2232} &  0.2570& 14:57:15.110& +22:20:32.26& 3.99   &  4192 &ACIS-I & 91886 \\
\multicolumn{2}{l}{MS 1512.4+3647} &  0.3726& 15:14:22.507& +36:36:20.15& 5.14   & 800 &ACIS-S & 14665 \\
\multicolumn{2}{l}{MS 1621.5+2640} &  0.4274& 16:23:35.37 & +26:34:19.4& 5.59  & 546&ACIS-I & 30062 \\
\enddata
  
\tablenotetext{a}{Corrected exposure time (see text for corrections applied).} 
\end{deluxetable}

\clearpage

\begin{deluxetable}{ccrrrrr}
\tablecolumns{7}
\tablewidth{0pc}
\tablecaption{$\beta$-Model Fits\label{table2}}
\tablehead{ 
\multicolumn{2}{c}{Cluster} & \colhead{$R_{\rm c}$ [$h_{70}^{-1} \rm{kpc}$]} &
\colhead{$\beta$}  & \colhead{$I_{\rm 0}$\tablenotemark{a}} &
\colhead{$I_{\rm B}$\tablenotemark{a}}  & \colhead{$\chi^2/\rm{DOF}$}
}
\startdata 
 \multicolumn{2}{l}{Abell 2390}& $63^{+2}_{-2}$ & $0.531_{- 0.003}^{+ 0.003}$ & $1575_{-35}^{+34}$ & $1.7_{-0.2}^{+0.2}$ & 368.8/145 \\ 
\multicolumn{2}{l}{MS 0015.9+1609}& $253^{+13}_{-13}$ & $0.72_{- 0.02}^{+ 0.02}$ & $100_{-4}^{+4}$ & $1.17_{-0.04}^{+0.04}$ & 316.2/230  \\ 
\multicolumn{2}{l}{MS 0302.7+1658}&   $27_{-4}^{+5}$ & $0.50_{- 0.02}^{+ 0.02}$ & $337_{-56}^{+67}$ & $0.89_{-0.04}^{+0.04}$ & 328.2/324  \\ 
\multicolumn{2}{l}{MS 0440.5+0204}&  $13_{-1}^{+2}$ & $0.444_{- 0.006}^{+ 0.006}$ & $933_{-93}^{+84}$ & $2.4_{-0.1}^{+0.1}$ & 372.2/191  \\ 
\multicolumn{2}{l}{MS 0451.5+0250}&  $520_{-43}^{+50}$ & $0.87_{- 0.08}^{+ 0.1}$ & $51_{-1}^{+1}$ & $1.7_{-0.3}^{+0.3}$ & 620.4/227 \\
\multicolumn{2}{l}{MS 0451.6-0305}&  $200^{+10}_{-10}$ & $0.73_{- 0.02}^{+ 0.02}$ & $154_{-6}^{+6}$ & $1.81_{-0.05}^{+0.05}$ & 257.3/205  \\ 
\multicolumn{2}{l}{MS 0839.8+2938}&  $58^{+3}_{-3}$ & $0.62_{- 0.01}^{+ 0.02}$ & $407_{-27}^{+37}$ & $2.33_{-0.07}^{+0.07}$ & 220.7/208 \\
\multicolumn{2}{l}{MS 0906.5+1110}&  $96_{-6}^{+7}$ & $0.60_{- 0.02}^{+ 0.02}$ & $185_{-7}^{+8}$ & $1.2_{-0.2}^{+0.2}$ & 177/174 \\ 
\multicolumn{2}{l}{MS 1006.0+1202}&  $129_{-13}^{+15}$ & $0.55_{- 0.02}^{+ 0.03}$ & $88_{-6}^{+7}$ & $0.8_{-0.1}^{+0.2}$ & 270.7/213  \\ 
\multicolumn{2}{l}{MS 1008.1-1224}&  $113^{+9}_{-9}$ & $0.55_{- 0.02}^{+ 0.02}$ & $101_{-6}^{+7}$ & $1.09_{-0.06}^{+0.06}$ & 329.4/218 \\ 
\multicolumn{2}{l}{MS 1358.4+6245}&  $76^{+5}_{-5}$ & $0.54_{- 0.01}^{+ 0.01}$ & $455_{-30}^{+32}$ & $3.8_{-0.1}^{+0.1}$ & 349.6/207  \\ 
\multicolumn{2}{l}{MS 1455.0+2232}&  $36.7^{+0.8}_{-0.8}$ & $0.612_{- 0.004}^{+ 0.004}$ & $3277_{-73}^{+78}$ & $2.29_{-0.06}^{+0.06}$ & 789.1/209  \\ 
\multicolumn{2}{l}{MS 1512.4+3647}&  $22^{+3}_{-3}$ & $0.51_{- 0.01}^{+ 0.01}$ & $688_{-88}^{+112}$ & $1.52_{-0.08}^{+0.08}$ & 192.5/191   \\ 
\multicolumn{2}{l}{MS 1621.5+2640}&  $184_{-25}^{+28}$ & $0.56_{- 0.04}^{+ 0.05}$ & $34_{-3}^{+3}$ & $1.38_{-0.1}^{+0.09}$ & 234/225  \\ 
\enddata
\tablenotetext{a}{Surface brightness $I$ in units of $10^{-9}$ photons sec${}^{-1}$ cm${}^{-2}$ arcsec${}^{-2}$}
\end{deluxetable}

\clearpage

\begin{deluxetable}{ccrrrrrrrr}\rotate
\tablecolumns{10}
\tablewidth{0pc}
\tablecaption{Double $\beta$-Model Fits\label{table3}}
\tablehead{ 
\multicolumn{2}{c}{Cluster} & \colhead{$R_{\rm c1}$} [$h_{70}^{-1} \rm{kpc}$] &
\colhead{$\beta_1$}  & \colhead{$I_{\rm 01}$\tablenotemark{a}} &
\colhead{$R_{\rm c2}$} [$h_{70}^{-1} \rm{kpc}$] & \colhead{$\beta_2$}  &
\colhead{$I_{\rm 02}$\tablenotemark{a}} & \colhead{$I_{\rm B}$\tablenotemark{a}}  & \colhead{$\chi^2/\rm{DOF}$}}
\startdata 
 \multicolumn{2}{l}{Abell 2390}& $20^{+4}_{-3}$ & $0.53_{- 0.07}^{+ 0.1}$ &  $1709_{-176}^{+157}$ &  $80_{-7}^{+11}$ & $0.52_{- 0.01}^{+ 0.02}$ &$504.9_{-84}^{+101}$  &  $1.7_{-0.2}^{+0.2}$ & 177.6/106  \\ 
 \multicolumn{2}{l}{Abell 0440.5+0204}& $11.7^{+1}_{-0.6}$ & $1.0_{- 0.2}^{+ 0.2}$ & $1246_{-134}^{+149}$ & $33_{-4}^{+2}$ & $0.489_{- 0.007}^{+ 0.011}$ & $338_{-17}^{+30}$ & $2.80_{-0.09}^{+0.1}$ & 221.5/191  \\ 
\multicolumn{2}{l}{MS 0839.8+2938}& $21^{+4}_{-4}$ & $0.9_{- 0.1}^{+ 0.2}$ & $761_{-105}^{+146}$ & $59_{-7}^{+8}$ & $0.62_{- 0.02}^{+ 0.02}$ & $366_{-61}^{+56}$ & $2.3_{-0.1}^{+0.1}$ & 190/167  \\ 
\multicolumn{2}{l}{MS 1358.4+6245}& $10_{-1}^{+2}$ & $0.46_{- 0.02}^{+ 0.05}$ & $3042_{-380}^{+423}$  &  $121_{-20}^{+24}$ & $0.63_{- 0.05}^{+ 0.1}$ & $188_{-31}^{+38}$  & $4.0_{-0.4}^{+0.3}$ & 156/139  \\ 
\multicolumn{2}{l}{MS 1455.0+2232}& $36_{-4}^{+8}$ & $1.1_{- 0.2}^{+ 0.2}$ & $2635_{-383}^{+325}$ & $56_{-18}^{+21}$ & $0.64_{- 0.01}^{+ 0.01}$ & $1306_{-218}^{+261}$ & $2.45_{-0.07}^{+0.07}$ &  276.3/183  \\ 
\enddata
\tablenotetext{a}{Surface brightness $I$ in units of $10^{-9}$ photons sec${}^{-1}$ cm${}^{-2}$ arcsec${}^{-2}$}
\end{deluxetable}

\clearpage

\begin{deluxetable}{ccccccc}
\tablecolumns{6}
\tablewidth{0pt}
\tablecaption{Single Temperature Spectral Fits for R $<$ $\rm{R}_{2500}$\label{table4}}
\tablehead{
\multicolumn{2}{c}{Cluster} & \colhead {$\rm{R}_{2500}$} & \colhead{kT} & \colhead{${\rm{N}}_{\rm{H}}$} &\colhead{Z} & \colhead{$\chi^2/\rm{DOF}$}  \\
\multicolumn{2}{c}{} & \colhead{[$\rm{h}_{70}^{-1}$ kpc]} & \colhead{[keV]} &\colhead{[${10}^{20}~{\rm{cm}}^{-2}$]} &\colhead{[solar]}& \colhead{}
}
\startdata
\multicolumn{2}{l}{Abell 2390} & {${701}^{+25}_{-19}$} & {${10.3}^{+0.6}_{-0.5}$} & 6.8 &    {${0.4}^{+0.1}_{-0.1}$}  &{401.8/434} \\
\multicolumn{2}{l}{MS 0015.9+1609} & {${550}^{+38}_{-35}$} & {${9.2}^{+1.0}_{-0.9}$} & 4.07 & {${0.3}^{+0.1}_{-0.1}$}  & {185.5/271} \\
\multicolumn{2}{l}{MS 0302.7+1658}  & {${352}^{+139}_{-72}$} & {${4}^{+3}_{-1}$} & 10.9 &	    {${0.1}^{+1}_{-0.1}$}  & {10.3/26} \\
\multicolumn{2}{l}{MS 0440.5+0204} &  {${640}^{+120}_{-68}$} & {${8}^{+2}_{-1}$} & 9.67 &    {${0.8}^{+0.6}_{-0.5}$}  & {272.2/378} \\
\multicolumn{2}{l}{} & {${583}^{+95}_{-66}$} & {${7}^{+2}_{-2}$} & {${12}^{+3}_{-3}$} &    {${0.8}^{+0.5}_{-0.4}$}  &{267.7/377} \\
\multicolumn{2}{l}{MS 0451.5+0250} & {${496}^{+90}_{-74}$} & {${6.1}^{+0.7}_{-0.6}$} & 7.8 & {${0.4}^{+0.1}_{-0.1}$}   &  {443.9/420} \\
\multicolumn{2}{l}{MS 0451.6-0305} & {${614}^{+48}_{-38}$} &{${10.2}^{+1}_{-1}$} & 5.07 & {${0.4}^{+0.1}_{-0.1}$}   &  {223/258} \\
\multicolumn{2}{l}{MS 0839.8+2938} & {${491}^{+24}_{-22}$} &{${4.0}^{+0.3}_{-0.3}$} & 4.11 &   {${0.6}^{+0.2}_{-0.2}$}   & {241.7/283} \\
\multicolumn{2}{l}{MS 0906.5+1110} & {${582}^{+44}_{-34}$} & {${5.6}^{+0.8}_{-0.6}$} & 3.54 &	    {${0.2}^{+0.2}_{-0.2}$}   & {134.2/319} \\
\multicolumn{2}{l}{MS 1006.0+1202} & {${557}^{+67}_{-45}$} &{${7}^{+1}_{-1}$} & 3.76 & {${0.3}^{+0.2}_{-0.2}$}   & {100/199} \\
\multicolumn{2}{l}{MS 1008.1-1224} & {${501}^{+41}_{-32}$} &{${5.9}^{+0.9}_{-0.7}$} & 6.98 &   {${0.2}^{+0.2}_{-0.2}$}   & {159.5/203} \\
\multicolumn{2}{l}{MS 1358.4+6245} & {${575}^{+72}_{-44}$} &{${8.0}^{+1.1}_{-0.9}$} & 1.93 &  {${0.5}^{+0.2}_{-0.2}$}   & {272.2/332}\\
\multicolumn{2}{l}{MS 1455.0+2232} & {${496}^{+7}_{-7}$} &{${4.4}^{+0.1}_{-0.1}$} & 3.13 &   {${0.38}^{+0.05}_{-0.05}$}  & {569.6/385} \\
\multicolumn{2}{l}{MS 1512.4+3647} & {${351}^{+42}_{-32}$} &{${3.4}^{+0.8}_{-0.7}$} & 1.38 &	    {${0.6}^{+0.5}_{-0.4}$}   & {43.2/74} \\
\multicolumn{2}{l}{MS 1621.5+2640} & {${493}^{+109}_{-68}$} & {${7}^{+3}_{-2}$} & 3.57 &	    {${0.4}^{+0.4}_{-0.4}$}  & {56.3/116} \\
\enddata
\end{deluxetable}

\clearpage

\begin{deluxetable}{cccccccc}
\tablecolumns{7}
\tablewidth{0pt}
\tablecaption{Cooling Core Corrected Single Temperature Spectral Fits for R $<$ $\rm{R}_{2500}$\label{table5}}
\tablehead{
\multicolumn{2}{c}{Cluster} & \colhead {$\rm{R}_{\rm{cutoff}}$} & \colhead {$\rm{R}_{2500}$} & \colhead{kT} & \colhead{${\rm{N}}_{\rm{H}}$} &\colhead{Z} & \colhead{$\chi^2$/DOF}  \\
\multicolumn{2}{c}{} & \colhead{[$\rm{h}_{70}^{-1}$ kpc]} & \colhead{[$\rm{h}_{70}^{-1}$ kpc]} & \colhead{[keV]} &\colhead{[${10}^{20}~{\rm{cm}}^{-2}$]} &\colhead{[solar]}& \colhead{}
}
\startdata
\multicolumn{2}{l}{Abell 2390} & {{66}} & {${761}^{+35}_{-30}$}& {${12.1}^{+1}_{-0.9}$}  & 6.8 &    {${0.3}^{+0.3}_{-0.3}$}  &{355/434} \\
\multicolumn{2}{l}{MS 0839.8+2938} & {${109}$} & {${565}^{+72}_{-55}$}&{${5}^{+1}_{-1}$} & 4.11 &   {${0.6}^{+0.5}_{-0.4}$}   & {195.5/237} \\
\multicolumn{2}{l}{MS 1358.4+6245} & {${154}$} & {${658}^{+93}_{-59}$}&{${9}^{+2}_{-1}$} & 1.93 &  {${0.4}^{+0.3}_{-0.3}$}   & {209/253}\\
\multicolumn{2}{l}{MS 1455.0+2232} & {${211}$} & {${575}^{+45}_{-35}$}&{${5.9}^{+0.9}_{-0.7}$} & 3.13 &   {${0.4}^{+0.2}_{-0.2}$}  & {238/273} \\
\enddata
\end{deluxetable}

\clearpage

\begin{deluxetable}{ccccc}
\tablecolumns{7}
\tablewidth{0pc}
\tablecaption{Mass Estimates for R $<$ $\rm{R}_{2500}$\label{table6a}}
\tablehead{                
\multicolumn{2}{c}{Cluster} & \colhead{${n}_0$} & \colhead{${\rm{M}}_{\rm{gas}}$} & \colhead{${\rm{M}}_{\rm{2500}}$}  \\
\multicolumn{2}{c}{}  & \colhead{[${10^{-1}~\rm{cm}^{-3}}$]} & \colhead{[$10^{13} ~\rm{h}_{70}^{-5/2} ~{\rm{M}_\odot}$]} & \colhead{[$10^{13} ~\rm{h}_{70}^{-1} ~{\rm{M}_\odot}$]} 
}
\startdata
\multicolumn{2}{l}{Abell 2390}   & $ 0.224_{-0.005}^{+ 0.005}$ & $  6.1_{ -0.4}^{+  0.3}$ & $ 59_{ -3}^{+  3}$  \\
\multicolumn{2}{l}{MS 0015.9+1609} & $ 0.0958_{-0.0008}^{+ 0.002}$ & $ 4.6_{ -0.1}^{+  0.1}$ & $38_{-3}^{+ 3}$  \\
 \multicolumn{2}{l}{MS 0302.7+1658}  & $ 0.46_{-0.01}^{+ 0.01}$ & $  0.8_{ -0.1}^{+  0.1}$ & $ 9_{-3}^{+ 3}$ \\
 \multicolumn{2}{l}{MS 0440.5+0204} & $ 0.280_{-0.004}^{+ 0.004}$ & $  1.62_{ -0.09}^{+  0.09}$ & $ 26_{-5}^{+ 4}$ \\
 \multicolumn{2}{l}{MS 0451.5+0250}  & $ 0.0341_{-0.0004}^{+ 0.0004}$ & $  2.8_{ -0.3}^{+  0.3}$ & $ 17.6_{ -0.9}^{+ 0.9}$\\
 \multicolumn{2}{l}{MS 0451.6-0305} & $ 0.1325_{-0.0009}^{+ 0.0009}$ & $ 5.2_{ -0.1}^{+  0.2}$ & $52_{-4}^{+ 4}$ \\
 \multicolumn{2}{l}{MS 0839.8+2938}   & $ 0.240_{-0.005}^{+ 0.005}$ & $  1.6_{ -0.1}^{+  0.1}$ & $ 23_{-3}^{+3}$ \\
 \multicolumn{2}{l}{MS 0906.5+1110}  & $ 0.123_{-0.001}^{+ 0.001}$ & $  2.00_{ -0.09}^{+  0.09}$ & $ 24_{-2}^{+ 2}$ \\
 \multicolumn{2}{l}{MS 1006.0+1202}   & $ 0.079_{-0.001}^{+ 0.001}$ & $ 2.3_{ -0.1}^{+  0.1}$ & $ 25_{-3}^{+ 3}$ \\
 \multicolumn{2}{l}{MS 1008.1-1224}  & $ 0.100_{-0.001}^{+ 0.001}$ & $  2.0_{ -0.1}^{+  0.1}$ & $ 19_{-2}^{+2}$ \\
 \multicolumn{2}{l}{MS 1358.4+6245}  & $ 0.139_{-0.005}^{+ 0.005}$ & $  3.4_{ -0.5}^{+ 0.5}$ & $47_{-6}^{+ 6}$ \\
 \multicolumn{2}{l}{MS 1455.0+2232} & $ 0.50_{-0.01}^{+ 0.01}$ & $  2.7_{ -0.3}^{+  0.3}$ & $  27_{ -2}^{+2}$\\
 \multicolumn{2}{l}{MS 1512.4+3647} & $ 0.65_{-0.01}^{+ 0.01}$ & $  0.86_{ -0.09}^{+  0.09}$ & $ 7_{-1}^{+1}$\\
 \multicolumn{2}{l}{MS 1621.5+2640}    & $ 0.052_{-0.001}^{+ 0.001}$ & $  1.9_{ -0.2}^{+  0.2}$ & $24_{-5}^{+5}$ \\
\enddata
\end{deluxetable}

\clearpage

\begin{deluxetable}{cccccc}
\tablecolumns{7}
\tablewidth{0pc}
\tablecaption{Mass Estimates for R $<$ $\rm{R}_{200}$\label{table6}}
\tablehead{                
\multicolumn{2}{c}{Cluster} & \colhead{$\rm{R}_{200}$\tablenotemark{a}} &  \colhead{${\rm{M}}_{\rm{gas}}$} & \colhead{${\rm{M}}_{\rm{200}}$} & \colhead{$f_{gas}$}  \\
\multicolumn{2}{c}{} & \colhead{[$\rm{h}_{70}^{-1}$ Mpc]}  & \colhead{[$10^{14} ~\rm{h}_{70}^{-5/2} ~{\rm{M}_\odot}$]} & \colhead{[$10^{14} ~\rm{h}_{70}^{-1} ~{\rm{M}_\odot}$]} & \colhead{[$h_{70}^{-3/2}$]} 
}
\startdata
\multicolumn{2}{l}{Abell 2390} & ${2.7}^{+0.1}_{-0.1}$  &  $  3.9_{ -0.3}^{+  0.3}$ & $ 21_{ -1}^{+  1}$ & $0.19^{+0.02}_{-0.02} $  \\
\multicolumn{2}{l}{MS 0015.9+1609}&  ${2.1}^{+0.1}_{-0.1}$ &  $ 2.62_{ -0.10}^{+  0.10}$ & $17_{-1}^{+ 1}$ &  $ 0.15^{+0.01}_{-0.01}$ \\
 \multicolumn{2}{l}{MS 0302.7+1658}&  ${1.2}^{+0.5}_{-0.3}$  & $  0.56_{ -0.07}^{+  0.07}$ & $ 3_{-1}^{+ 1}$ &  $ 0.17^{+0.06}_{-0.06}$\\
 \multicolumn{2}{l}{MS 0440.5+0204}&  ${2.1}^{+0.3}_{-0.2}$ &  $  1.14_{ -0.07}^{+  0.07}$ & $ 9_{-2}^{+ 2}$ &  $ 0.13^{+0.02}_{-0.02}$ \\
 \multicolumn{2}{l}{MS 0451.5+0250}&  ${2.5}^{+0.2}_{-0.1}$  & $  2.8_{ -0.3}^{+  0.3}$ & $ 15.9_{ -0.8}^{+ 0.8}$ & $0.17^{+0.02}_{-0.02}$\\
 \multicolumn{2}{l}{MS 0451.6-0305}& ${2.3}^{+0.2}_{-0.1}$ &  $ 2.33_{ -0.10}^{+  0.10}$ & $21_{-2}^{+ 2}$ &  $ 0.11^{+0.01}_{-0.01}$\\
 \multicolumn{2}{l}{MS 0839.8+2938}& ${2.0}^{+0.3}_{-0.2}$   &  $  0.73_{ -0.06}^{+  0.06}$ & $ 8_{-1}^{+ 1}$ &  $ 0.09^{+0.01}_{-0.02}$\\
 \multicolumn{2}{l}{MS 0906.5+1110}& ${2.1}^{+0.2}_{-0.1}$   & $  1.05_{ -0.05}^{+  0.05}$ & $ 8.6_{-0.7}^{+ 0.7}$ &  $ 0.12^{+0.01}_{-0.01}$\\
 \multicolumn{2}{l}{MS 1006.0+1202}& ${2.0}^{+0.2}_{-0.2}$    & $ 1.6_{ -0.1}^{+  0.1}$ & $ 9_{-1}^{+ 1}$ &  $ 0.17^{+0.02}_{-0.02}$\\
 \multicolumn{2}{l}{MS 1008.1-1224}& ${1.8}^{+0.1}_{-0.1}$  & $  1.34_{ -0.07}^{+  0.07}$ & $ 7.2_{-0.6}^{+0.6}$ &  $ 0.19^{+0.02}_{-0.02}$\\
 \multicolumn{2}{l}{MS 1358.4+6245}& ${2.4}^{+0.3}_{-0.2}$   & $  1.7_{ -0.3}^{+ 0.3}$ & $17_{-2}^{+ 2}$ &  $ 0.09^{+0.02}_{-0.02}$\\
 \multicolumn{2}{l}{MS 1455.0+2232}& ${2.04}^{+0.2}_{-0.1}$  & $  1.2_{ -0.1}^{+  0.1}$ & $  9.5_{ -0.9}^{+  0.9}$ & $0.12^{+0.02}_{-0.02}$ \\
 \multicolumn{2}{l}{MS 1512.4+3647}& ${1.3}^{+0.1}_{-0.1}$   & $  0.59_{ -0.04}^{+  0.04}$ & $ 2.7_{ -0.4}^{+  0.3}$ &  $ 0.22^{+0.03}_{-0.03}$\\
 \multicolumn{2}{l}{MS 1621.5+2640}& ${1.8}^{+0.4}_{-0.2}$   & $  1.5_{ -0.1}^{+  0.1}$ & $10_{-2}^{+ 2}$ &  $ 0.15^{+0.03}_{-0.03}$\\
\enddata
\tablenotetext{a}{90\% confidence intervals}
\end{deluxetable}

\clearpage

\begin{deluxetable}{ccccc}
\tablecolumns{5}
\tablewidth{0pc}
\tablecaption{Dynamical Mass Comparisons for R $<$ $\rm{R}_{200}$\label{table7}}
\tablehead{                
\multicolumn{2}{c}{Cluster} & \colhead{Dynamical Mass} & \colhead{X-ray Mass} & \colhead{Ratio}   \\
\multicolumn{2}{c}{} & \colhead{[$10^{14} ~\rm{h}_{70}^{-1} ~{\rm{M}_\odot}$]} & \colhead{[$10^{14} ~\rm{h}_{70}^{-1} ~{\rm{M}_\odot}$]} &  \colhead{[Dynamical/X-ray]} 
}
\startdata
\multicolumn{2}{l}{Abell 2390} & ${26}^{+4}_{-3}$  & $21_{-1}^{+ 1}$  & $ 1.2_{-0.2}^{+0.2}$ \\
\multicolumn{2}{l}{MS 0015.9+1609}&  ${16}^{+5}_{-3}$  & $17_{-1}^{+ 1}$ & $ 0.9_{-0.2}^{+0.3}$ \\
 \multicolumn{2}{l}{MS 0302.7+1658}& ${3.8}^{+1}_{-0.6}$  &$ 3_{-1}^{+ 1}$  & $  1.3_{-0.5}^{+0.5}$ \\
 \multicolumn{2}{l}{MS 0440.5+0204}& ${6}^{+2}_{-1}$  & $ 9_{-2}^{+ 2}$ & $  0.7_{-0.2}^{+0.3}$ \\
 \multicolumn{2}{l}{MS 0451.5+0250}& ${15}^{+2}_{-2}$  & $15.9_{-0.8}^{+0.8}$ & $  0.9_{-0.1}^{+0.1}$ \\
 \multicolumn{2}{l}{MS 0451.6-0305}& ${24}^{+4}_{-3}$  & $21_{-2}^{+ 2}$ & $  1.1_{-0.2}^{+0.2}$ \\
 \multicolumn{2}{l}{MS 0839.8+2938}& ${12}^{+3}_{-3}$  & $ 8_{-1}^{+ 1}$ & $  1.5_{ -0.4}^{+0.4}$ \\
 \multicolumn{2}{l}{MS 0906.5+1110}& ${8}^{+2}_{-1}$  & $ 8.6_{-0.7}^{+ 0.7}$ & $  0.9_{ -0.1}^{+ 0.2 }$ \\
 \multicolumn{2}{l}{MS 1006.0+1202}&  ${13}^{+4}_{-3}$  &  $ 9_{-1}^{+1}$ & $  1.4_{ -0.4}^{+ 0.5 }$ \\
 \multicolumn{2}{l}{MS 1008.1-1224}& ${12}^{+3}_{-2}$  &  $ 7.2_{-0.6}^{+ 0.6}$ & $  1.7_{ -0.3}^{+0.4  }$\\
 \multicolumn{2}{l}{MS 1358.4+6245}& ${14}^{+2}_{-1}$  & $17_{-2}^{+ 2}$ & $  0.8_{-0.1}^{+0.2 }$ \\
 \multicolumn{2}{l}{MS 1455.0+2232}&${12}^{+3}_{-2}$  & $9.5_{ -0.9}^{+0.9 }$ & $  1.3_{ -0.2}^{+0.3}$ \\
 \multicolumn{2}{l}{MS 1512.4+3647}& ${2.6}^{+1}_{-0.8}$  & $ 2.7_{ -0.4}^{+  0.3}$ & $  1.0_{ -0.3}^{+ 0.4}$\\
 \multicolumn{2}{l}{MS 1621.5+2640}&  ${8}^{+1}_{-1}$  &   $10_{-2}^{+ 2}$ & $  0.8_{ -0.2}^{+0.2}$  \\
\enddata
\end{deluxetable}

\clearpage

\begin{deluxetable}{ccccc}
\tablecolumns{5}
\tablewidth{0pc}
\tablecaption{Weak Lensing Mass Comparisons for R $<$ 500 $\rm{h}^{-1}$ kpc\label{table8}}
\tablehead{                
\multicolumn{2}{c}{Cluster} & \colhead{Weak Lensing Mass}  &\colhead{X-ray Mass} & \colhead{Ratio} \\
\multicolumn{2}{c}{} &  \colhead{[$10^{14} ~\rm{h}^{-1} ~{\rm{M}_\odot}$]}  & \colhead{[$10^{14} ~\rm{h}^{-1} ~{\rm{M}_\odot}$]} & \colhead{[Lensing/X-ray]}
}
\startdata
\multicolumn{2}{l}{Abell 2390} & ${5.0}^{+0.6}_{-0.6}$  & $5.3_{-0.4}^{+0.5 }$ & ${0.9}^{+0.1}_{-0.1}$  \\
\multicolumn{2}{l}{MS 0015.9+1609}&  ${8}^{+1}_{-1}$  & $ 5.2_{ -0.4}^{+  0.4}$ & ${1.5}^{+0.3}_{-0.3}$  \\
 \multicolumn{2}{l}{MS 0906.5+1110}& ${3.5}^{+0.7}_{-0.7}$  & $ 2.8_{ -0.2}^{+  0.2}$ & ${1.2}^{+0.3}_{-0.3}$  \\
 \multicolumn{2}{l}{MS 1358.4+6245}& ${4.2}^{+0.8}_{-0.8}$   & $ 4.9_{-0.6}^{+ 0.6}$ & ${0.9}^{+0.2}_{-0.2}$  \\
 \multicolumn{2}{l}{MS 1455.0+2232}&${3.2}^{+0.7}_{-0.7}$  &  $ 3.2_{-0.3}^{+0.3}$ & ${1.0}^{+0.2}_{-0.2}$  \\
 \multicolumn{2}{l}{MS 1512.4+3647}& ${2.0}^{+0.8}_{-0.8}$   &  $ 1.4_{ -0.2}^{+  0.2}$& ${1.4}^{+0.6}_{-0.6}$  \\
 \multicolumn{2}{l}{MS 1621.5+2640}&  ${5}^{+1}_{-1}$   & $ 3.6_{-0.7}^{+ 0.7}$  & ${1.4}^{+0.4}_{-0.4}$  \\
\enddata
\end{deluxetable}

\clearpage

\begin{deluxetable}{cccc}
\tablecolumns{4}
\tablewidth{0pt}
\tablecaption{Cluster Richness and Luminosity\label{table9}}
\tablehead{
\multicolumn{2}{c}{Cluster} & 
\colhead{${\rm{L}}_{\rm{x}}$\tablenotemark{a}} & 
\colhead{${\rm{Bgc}_{\rm{red}}}$}  \\
\multicolumn{2}{c}{} & 
\colhead{[$10^{44}~{\rm{erg}}~{\rm{s}}^{-1}$]} & 
\colhead{[$\rm{Mpc}^{1.77}$]}  
}
\startdata
\multicolumn{2}{l}{Abell 2390} &  $56.1^{+0.2}_{-0.2}$\tablenotemark{b} & $1473\pm{218}$ \\
\multicolumn{2}{l}{MS 0015.9+1609} &   $41.0^{+0.7}_{-1}$ & $1789\pm{260}$ \\
\multicolumn{2}{l}{MS 0302.7+1658} &   $5^{+1}_{-1}$ &$762\pm{201}$  \\
\multicolumn{2}{l}{MS 0440.5+0204} &   $6.2^{+0.2}_{-0.2}$ &$559\pm{154}$ \\
\multicolumn{2}{l}{MS 0451.5+0250} &   $15.0^{+0.3}_{-0.5}$ & $1198\pm{208}$   \\
\multicolumn{2}{l}{MS 0451.6-0305} &   $49.1^{+0.9}_{-1.}$ & $1232\pm{180}$   \\
\multicolumn{2}{l}{MS 0839.8+2938} &   $5.97^{+0.1}_{-0.08}$\tablenotemark{b}  & $1106\pm{187}$ \\
\multicolumn{2}{l}{MS 0906.5+1110} &   $8.5^{+0.3}_{-0.3}$ & $1134\pm{209}$  \\
\multicolumn{2}{l}{MS 1006.0+1202} &  $ 9.2^{+0.6}_{-0.6}$ & $1330\pm{216}$  \\
\multicolumn{2}{l}{MS 1008.1-1224} &   $9.0^{+0.3}_{-0.5}$ & $1444\pm{212}$  \\
\multicolumn{2}{l}{MS 1358.4+6245} &   $17.7^{+0.2}_{-0.2}$\tablenotemark{b} & $1182\pm{205}$   \\
\multicolumn{2}{l}{MS 1455.0+2232} &   $23.2^{+0.1}_{-0.2}$\tablenotemark{b} & $518\pm{154}$   \\
\multicolumn{2}{l}{MS 1512.4+3647} &  $ 4.9^{+0.4}_{-0.8}$ & $582\pm{180}$   \\
\multicolumn{2}{l}{MS 1621.5+2640} &   $8.7^{+0.5}_{-0.8}$ & $996\pm{199}$  \\
\enddata

\tablenotetext{a}{Unabsorbed bolometric X-ray luminosity for $\rm{R}<\rm{R}_{2500}$}
\tablenotetext{b}{Cooling core corrected}

\end{deluxetable}

\clearpage

\begin{deluxetable}{ccc}
\tablecolumns{3}
\tablewidth{0pt}
\tablecaption{Fitting Parameters ($\Delta=2500$) \label{table10}}
\tablehead{
\colhead{Relationship} &
\colhead{$C_1$} & 
\colhead{$C_2$}
}
\startdata
$T_{X}-B_{gc,red}$ & $-2.4\pm{0.8}$ & $1.1\pm{0.3}$ \\
${E(z)}^{-1}L_{X}-B_{gc,red}$ & $-6\pm{2}$ & $2.3\pm{0.7}$ \\
${E(z)}~\rm{M}_{2500}-B_{gc,red}$ &$9\pm{2}$ &$1.8\pm{0.5}$ \\
\enddata
\end{deluxetable}


\begin{thebibliography}{}

\bibitem[Abell(1958)]{abell} Abell, G.O. 1958, \apj, 3, 221
\bibitem[Akritas \& Bershady(1996)]{akritas} Akritas, M.G. \& Bershady, M.A. 1996, \apj, 470, 706
\bibitem[Allen \& Fabian(1997)]{allen97} Allen, S.W. \& Fabian, A.C. 1997, MNRAS, 286, 583
\bibitem[Allen(2000)]{allen00} Allen, S.W. 2000, MNRAS, 315, 269
\bibitem[Allen \etal(2002)]{allen02} Allen, S.W., Schmidt, R.W. \& Fabian, A.C. 2002, MNRAS, 334, 11
\bibitem[Allen, Schmidt, \& Fabian(2001)]{allen01} Allen, S.W., Schmidt, R.W. \& Fabian, A.C. 2001, MNRAS 328, L37
\bibitem[Arabadjis, Bautz, \& Garmire(2002)]{arabadjis} Arabadjis, J.S., Bautz, M.W., \& Garmire, G.P. 2002, \apj, 572, 66
\bibitem[Arnaud(1996)]{arnaud96} Arnaud, K.A.  1996, ADASS, 101, 5
\bibitem[Arnaud \& Evrard(1999)]{arnaud99} Arnaud, M, Evrard, A.E. 1999, MNRAS 305, 631
\bibitem[Bahcall, Fan, \& Cen(1997)]{bahcall} Bahcall, N.A., Fan, X., and Cen, R. 1997, \apj, 485, 53
\bibitem[Balland \& Blanchard(1997)]{balland} Balland, C. \& Blanchard, A. 1997, \apj, 487, 33
\bibitem[Bird, Mushotzky, \& Metzler(1995)]{bird} Bird, C.M., Mushotzky, R.F., \& Metzler, C.A. 1995, \apj, 453, 40
\bibitem[B{\"o}hringer \etal(1998)]{bohringer} B{\"o}hringer, H., Tanaka, Y., Mushotkzy, R.F., Ikebe, Y. \& Hattori, M. 1998, A\&A, 334, 789
\bibitem[Borgani \etal(1999)]{borgani} Borgani, S., Girardi, M., Carlberg, R.G., Yee, H.K.C., \& Ellingson, E. 1999, \apj, 527, 561
\bibitem[Carlberg \etal(1996)]{carlberg} Carlberg, R.G., Yee, H.K.C., Ellingson, E., Abraham, R., Gravel, P., Morris, S., \& Pritchet, C.J. 1996, \apj, 462, 32
\bibitem[Cen(1997)]{cen} Cen, R. 1997, \apj, 485, 39
\bibitem[Davis \& Peebles(1983)]{davis} Davis, M. \& Peebles, P.J.E. 1983, /apj, 294, 70
\bibitem[Dickey \& Lockman(1990)]{dickey90} Dickey, J.M. \& Lockman, F.J.  1990, ARA\&A, 28, 215
\bibitem[Donahue, Stocke, \& Gioia(1992)]{donahue92} Donahue, M., Stocke, J.T., \& Gioia, I.M. 1992, \apj, 385, 49
\bibitem[Donahue \etal(2003)]{donahue03} Donahue, M., Gaskin, J.A., Patel, S.K., Joy, M., Clowe, D., \& Hughes, J.P. 2003, \apj, 598, 190
\bibitem[Eke et al.(1998)]{eke} Eke, V.R., Cole, S., Frenk, C.S., \& Henry, J.P. 1998, MNRAS, 298, 1145
\bibitem[Ellingson \etal(2001)]{ellingson} Ellingson, E., Lin, H., Yee, H.K.C., \& Carlberg, R.G. 2001, \apj, 547, 609
\bibitem[Ettori(2000)]{ettori00} Ettori, S. 2000, MNRAS, 311, 313
\bibitem[Ettori, De Grandi, \& Molendi(2002)]{ettori02} Ettori, S., De Grandi, S., \& Molendi, S. 2002, A\&A, 391, 841
\bibitem[Ettori \& Lombardi(2003)]{ettori03} Ettori, S., \& Lombardi, M. 2003, A\&A, 398, 5
\bibitem[Ettori \etal(2004)]{ettori04} Ettori, S., Tozzi, P., Borgani, S, \& Rosati, P. 2004, A\&A, 417, 13
\bibitem[Fukugita, Hogan, \& Peebles(1998)]{fukugita} Fukugita, M., Hogan, C.J., \& Peebles, P.J.E. 1998, \apj, 503, 518
\bibitem[Gioia \etal(1990)]{gioia90} Gioia, I.M., Henry, J.P., Maccacaro, T., Morris, S.L., Stocke, J.T., \& Wolter, A.  1990, \apj, 356, L35 
\bibitem[Gladders \& Yee(2005)]{gladders05} Gladders, M.D. \& Yee, H.K.C. 2005, ApJS, 157, 1
\bibitem[Henry \etal(1992)]{henry92} Henry, J.P., Gioia, I.M., Maccacaro, T., Morris, S.L., Stocke, J.T., \& Wolter, A.  1992, \apj, 386, 408 
\bibitem[Henry(2004)]{henry04} Henry, J.P. 2004, \apj, 609, 603
\bibitem[Hoekstra(2001)]{hoekstra01} Hoekstra, H. 2001, A\&A, 370, 743
\bibitem[Hoekstra(2006)]{hoekstra} Hoekstra, H., in preparation
\bibitem[Levine, Schulz, \& White(2002)]{levine02} Levine, E.S., Schulz, A.E., \& White, M.  2002, \apj, 577, 569
\bibitem[Lewis \etal(1999)]{lewis} Lewis, A.D., Ellingson, E., Morris, S.L., \& Carlberg, R.G. 1999, \apj, 517, 587
\bibitem[Longair \& Seldner(1979)]{long} Longair, M.S. \& Seldner, M. 1979, MNRAS, 189, 433
\bibitem[Matsumoto \etal(2000)]{matsumoto} Matsumoto, H., Tsuru, T.G., Fukazawa, Y., Hattoir, M., \& Davis, D.S. 2000, PASJ, 52, 153
\bibitem[Metzler, White, \& Loken(2001)]{metzler01} Metzler, C.A., White, M., \& Loken, C. \apj, 2001, 547, 560
\bibitem[Mohr \etal(2000)]{mohr} Mohr, J.J., Reese, E.D., Ellingson, E., Lewis, A.D., \& Evrard, A.E. 2000, \apj, 544, 109
\bibitem[Mushotzky \& Loewenstein(1997)]{richard} Mushotzky, R.F. \& Loewenstein 1997, \apj, 481, 63
\bibitem[Neumann \& Bohringer(1997)]{neumann} Neumann, D.M. \& Bohringer, H. 1997, MNRAS, 289, 123
\bibitem[Spergel \etal(2007)]{spergel2} Spergel, D.N. \etal 2007, ApJS, 170, 377
\bibitem[Van der Marel \etal(2000)]{vandermarel} Van der Marel, R.P., Magorrian, J., Carlberg, R.G, Yee, H.K.C., \& Ellingson, E. 2000, AJ, 119, 2038
\bibitem[van Dokkum et al.(1998)]{vandokkum} van Dokkum, P.G., Franx, M., Kelson, D.D. \& Illingworth, G.D. 1998, \apjl, 504, L17 
\bibitem[Voit(2004)]{voit} Voit, G.M. 2004, astroph/0410173
\bibitem[White \etal(1993)]{white93} White, S.D.M., Navarro, J.F., Evrard, A.E., \& Frenk, C.S. 1993, Nature, 366, 429
\bibitem[Yee, Ellingson, \& Carlberg(1996)]{yee96} Yee, H.K.C., Ellingson, E., \& Carlberg, R.G.  1996, \apjs, 102, 269
\bibitem[Yee \etal(1996)]{cnoccat} Yee, H.K.C., Ellingson, E., Abraham, R.G., Gravel, P., Carlberg, R.G., Smecker-Hane, T.A., Schade, D., \& Rigler, M. 1996, \apjs, 102, 289
\bibitem[Yee and Lopez-Cruz(1999)]{yee99} Yee, H.K.C., Lopez-Cruz, O. 1999, AJ, 117, 1985
\bibitem[Yee and Ellingson(2003)]{yee03} Yee, H.K.C., Ellingson, E. 2003, \apj, 585, 215

\end{thebibliography}
\end{document}